\documentclass[%
 preprint,
 amsmath,amssymb,
 aps,
]{revtex4-1}

\usepackage{graphicx}
\graphicspath{{figure/}}
\usepackage{overpic}
\usepackage{dcolumn}
\usepackage{bm}
\usepackage{float}
\usepackage{algorithm}
\usepackage{algorithmic}
\usepackage{setspace}

\begin{document}

\title{The Optimal Beam-loading\\
in Two-bunch Nonlinear Plasma Wakefield Accelerators}

\author{Xiaoning Wang\textsuperscript{1,2}}
\author{Jie Gao\textsuperscript{1,2}}
\author{Qianqian Su\textsuperscript{3}}
\author{Jia Wang\textsuperscript{1,2}}
\author{Dazhang Li\textsuperscript{1,2}}
\author{Ming Zeng\textsuperscript{1,2}}
\author{Wei Lu\textsuperscript{4}}
\author{Warren B.\ Mori\textsuperscript{3}}
\author{Chan Joshi\textsuperscript{3}}
\author{Weiming An\textsuperscript{5}}
\email[Author of correspondence: ]{anweiming@bnu.edu.cn}
\affiliation{\small{{\rm\textsuperscript{1}}Institute of High Energy Physics,
Chinese Academy of Sciences, Beijing 100049, China\\
{\rm\textsuperscript{2}}University of Chinese Academy of Sciences, Beijing 100049, China\\
{\rm\textsuperscript{3}}University of California Los Angeles, Los Angeles, California 90095, USA\\
{\rm\textsuperscript{4}}Department of Engineering Physics, Tsinghua University, Beijing 100084, China\\
{\rm\textsuperscript{5}}Department of Astronomy, Beijing Normal University, Beijing 100875, China}}

\begin{spacing}{1.2}
\begin{abstract}
Due to the highly nonlinear nature of the beam-loading, it is at present not possible to analytically determine the beam parameters needed in a two-bunch plasma wakefield accelerator for maintaining a low energy spread. Therefore in this paper, by using the Broyden-Fletcher–Goldfarb–Shanno algorithm for the parameter scanning with the code QuickPIC and the polynomial regression together with $k$-fold cross-validation method, we obtain two fitting formulas for calculating the parameters of tri-Gaussian electron beams when minimizing the energy spread based on the beam-loading effect in a nonlinear plasma wakefield accelerator. One formula allows the optimization of the normalized charge per unit length of a trailing beam to achieve the minimal energy spread, i.e. the optimal beam-loading. The other one directly gives the transformer ratio when the trailing beam achieves the optimal beam-loading. A simple scaling law for charges of drive beams and trailing beams is obtained from the fitting formula, which indicates that the optimal beam-loading is always achieved for a given charge ratio of the two beams when the length and separation of two beams and the plasma density are fixed. The formulas can also help obtain the optimal plasma densities for the maximum accelerated charge and the maximum acceleration efficiency under the optimal beam-loading respectively. These two fitting formulas will significantly enhance the efficiency for designing and optimizing a two-bunch plasma wakefield acceleration stage.
\end{abstract}
\end{spacing}


\maketitle

\section{Introduction}

Plasma-based acceleration (PBA) uses an intense laser pulse~\cite{RN88} or a charged particle beam~\cite{RN8} to excite a plasma wake, which can be utilized to accelerate electrons and positrons with high acceleration gradients~\cite{RNS1,RNS2,RN131,RN11,RN13,RNS3}. The acceleration gradients inside the plasma wake can easily exceed $10\ \rm{GeV}/m$~\cite{RN111,RN113,RN114,RN131,RNS3}, which are orders of magnitude higher than that of conventional accelerators. Such high acceleration gradients can significantly reduce the size and the cost of accelerators. This makes the PBA a promising candidate for the future linear colliders or light sources. Recently, the particle-beam-driven plasma wakefield acceleration (PWFA) has attracted a lot of attention due to tremendous theoretical and experimental progress~\cite{RN122,RN123,RN121,RN120,RN80,RN82,RN131,RN152,RN11,RN13,RN76,RN177,RNEp1}. There have been increasing numbers of facilities that are built for conducting PWFA research, such as Facilities for Accelerator Science and Experimental Test (FACET) II~\cite{RN14}, Advanced Proton Driven Plasma Wakefield Acceleration Experiment (AWAKE)~\cite{RN204}, Future Oriented Wakefield Accelerator Research and Development at FLASH (FLASHForward)~\cite{RN136} and EuPRAXIA~\cite{RN203}. In PWFA, when the highly relativistic drive beam passes through the plasma and its self-field is intense enough to expel all the plasma electrons away from the axis, a plasma bubble filled with plasma ions can be formed and moves along with the drive beam (which is the so-called blowout regime)~\cite{RN80}. As a result, the trailing beam will continuously gain energy until the drive beam exhausts its energy and no longer excites the plasma bubble.

In the blowout regime, when the trailing beam is loaded into the plasma wake, the longitudinal electric field of the wake will be modified. When the trailing beam is properly loaded (optimal beam-loading~\cite{RN152}), the longitudinal electric field felt by the trailing beam is locally flattened so that all the contained particles can be accelerated at the same rate resulting in the smallest increase in the energy spread as required by most accelerator applications. This beam-loading effect plays an important role on the beam quality and has been actively studied~\cite{RNN76,RN75,RN152,RNmulti,RNEp1,RNEp2,RNEp3}. Scaling laws for beam-loading are always useful as the guidance to design the PBA stage efficiently. There were two scaling laws proposed for a laser-driven stage. The number of particles loaded into a 3D bubble wake excited by a laser driver was found to scale with the normalized volume of the bubble or the square root of the laser power~\cite{RN75}. A similar scaling law but with a distinct parameter space was also offered by Ref.~[\onlinecite{RNN76}]. However, these scaling laws did not give the exact coefficient and the proper place for loading the trailing beam. In Ref.~[\onlinecite{RN152}], an analytical theory was proposed for beam-loading effect in the blowout regime to maintain the energy spread of the trailing beam. The charge, the shape and the placing of the trailing beam can be estimated for both a laser-driven stage and a beam-driven stage via this theory. However, when designing a two-bunch PWFA stage, the theory provided by Ref.~[\onlinecite{RN152}] is still not easy to use because it lacks the parameters for the drive beam. In addition, this analytical theory was obtained based on the assumption that the maximal normalized bubble radius is much larger than 1. Due to the limitation on the beam peak currents at present PWFA facilities, most PWFA experiments are conducted at a smaller maximal bubble radius, and no analytical model exists to predict their performances. Therefore, we here take a numerical approach to provide fitting formulas for the optimal beam-loading in a data-driven way that will help the design of two-bunch PWFA experiments. The fitting formulas consider parameters for both drive beam and trailing beam. In Sec.~\ref{opt}, the method to find the optimal beam-loading in a two-bunch PWFA stage is discussed. Subsequently, two fitting formulas are given in Sec.~\ref{res}. Specifically speaking, their availability for trailing beams with a longitudinal flat-top profile or a longitudinal trapezoidal profile are discussed in Sec.~\ref{extra}. In Sec.~\ref{charge}, the scaling law for charges of drive beams and trailing beams under the optimal beam-loading is derived from the fitting formulator. In Sec.~\ref{eng}, the optimal plasma densities for the maximum accelerated charge and maximum acceleration efficiency under the optimal beam-loading are discussed. In the last section, we summarize the results presented in this paper.

\section{Two-Bunch PWFA with Optimal Beam-loading}\label{opt}

\subsection{Optimization of Beam Parameters}
In a two-bunch PWFA stage, when the blowout occurs, the beam energy spread is mainly affected by the longitudinal wakefield~\cite{RN80}. Thus, having the longitudinal wakefield within the trailing beam as flat as possible is the most effective method to preserve beam energy spread. Parameters including beam charge $Q$, rms beam length $\sigma_z$, rms beam spot size $\sigma_r$, beam separation $d$ and plasma density $n_p$ are usually considered in a two-bunch PWFA design. For tri-Gaussian beams, the beam separation is defined as the distance between the center of the drive beam and that of the trailing beam. Electron beams with a tri-Gaussian profile have $\rho_b = n_b\cdot\exp(-\frac{x^2+y^2}{2\sigma_r^2})\exp(-\frac{\xi^2}{2\sigma_z^2})$, where $\xi=ct-z$ is the co-moving coordinate, $x$ and $y$ are the transverse coordinates, and the beam peak density is $n_b = \frac{N_b}{(2\pi)^{3/2}\sigma_z\sigma_r^2}$ where $N_b$ is the total number of electrons in the beam~\cite{RN199}. In this paper, we adopt normalized units. The beam density is normalized to the plasma density $n_p$ and the charge density is normalized to $en_p$ where $e$ is the electron charge. The length is normalized to the plasma skin depth $k_p^{-1} \equiv c/\omega_p$, where $c$ is the speed of light and $\omega_p = \sqrt{4\pi e^2n_p/m_e}$ is the plasma frequency where $m_e$ is the electron mass. The electric field is normalized to $m_ec\omega_p/e$. By using normalized units, we can drop the dependency of plasma density to simplify the model. Actually, engineering formulas that take the plasma density into account (described in Sec.~\ref{eng}) can be easily obtained from our fitting formulas in normalized units.
\vspace{0.2cm}

\begin{figure*}[htb]
    \centering
    \begin{overpic}
        [width=0.5\textwidth,height=5cm,trim=0 0 50 20,clip]{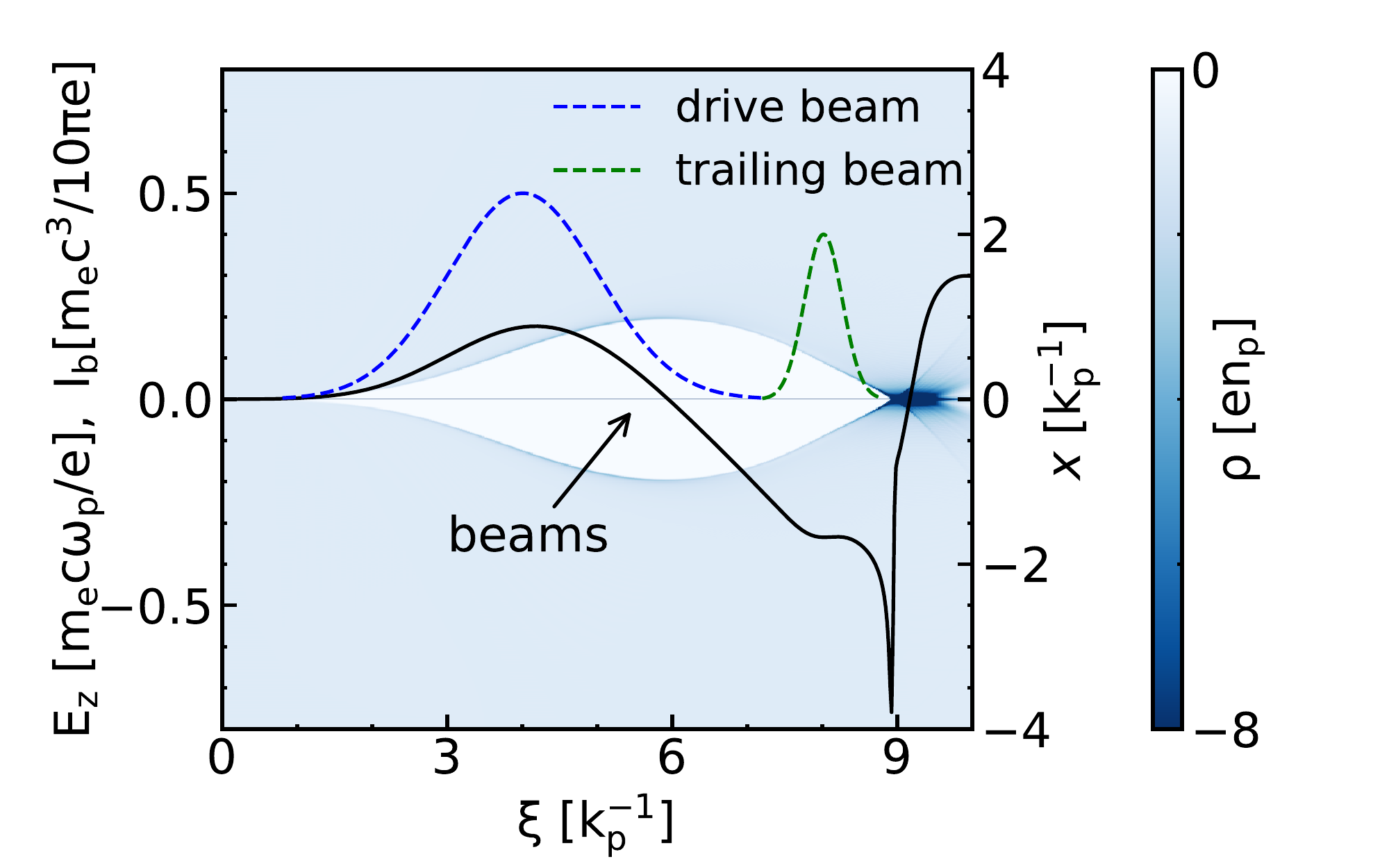}
        \put(18,54){\small{(a)}}
    \end{overpic}
    \begin{overpic}
    [width=0.4\textwidth,height=5cm,trim=0 0 45 20,clip]{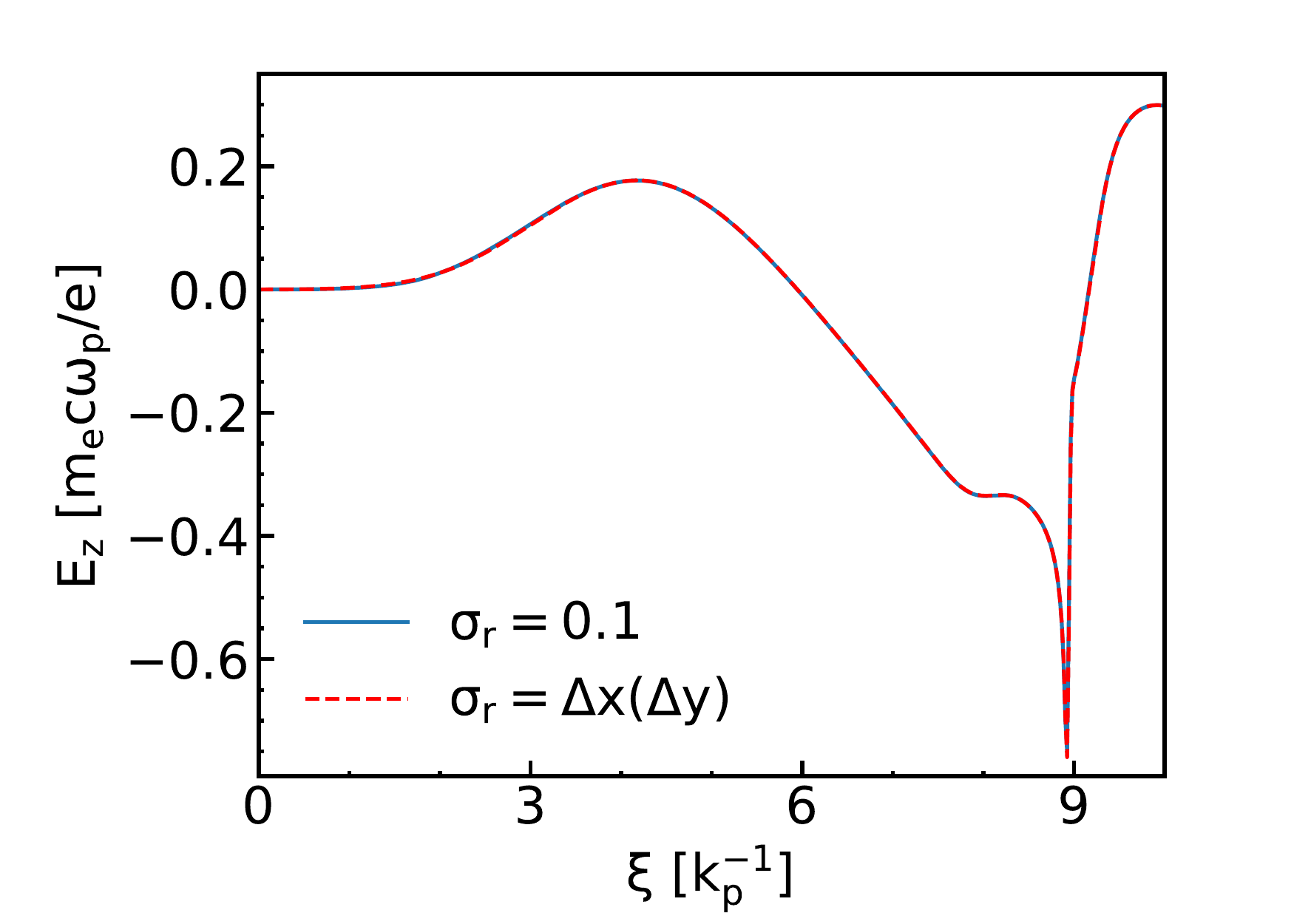}
    \put(23,68){\small{(b)}}
    \end{overpic}
\renewcommand{\baselinestretch}{1.5}  
\vspace{-0.8cm}
\caption{(a) QuickPIC simulation results of the plasma density and the density of one cell size drive beam and trailing beam. The solid line is the on-axis $E_z$ of the wakefield. The blue and green dashed lines represent the current ($I_b$) of the drive and trailing beam, respectively. The unit of $I_b$ is two-fifths of the normalized unit $m_ec^3/4\pi e\simeq8.5 {\rm kA}$. (b) The on-axis $E_z$ from the wakefield driven by one cell wide beams and beams with $\sigma_r = 0.1$. \label{f1}}
\vspace{-0.2cm}
\end{figure*}

In the blowout regime, if the bubble radius $R_b$ is much larger than the rms beam spot size $\sigma_r$, any variation within the beam spot size for the same charge per unit length $\Lambda = n_b\sigma^2_r$ will hardly change the wake~\cite{RN210}. In other words, the acceleration structure is determined by $\Lambda$ as long as $R_b\gg\sigma_r$ and the beam length is fixed. Therefore, we assume the beam has a very small spot size like a delta-function, in which case the dependency of the beam spot size is neglected. The delta-function-like beam is implemented in the simulation code QuickPIC~\cite{RN72,RN72web} by modifying the subroutine to directly initialize the beam density on the axis, which indicates that the beam has a spot size equal to the transverse cell size as shown in Fig.~\ref{f1} (a). In this simulation, the simulation box has the size of $8.0\times 8.0\times 10.0$ ($x,y,\xi$) with $512\times 512\times 512$ cells. The drive beam has $\Lambda_d = 0.2$ while the trailing beam has $\Lambda_t = 0.16$. The length of the drive beam and that of the trailing beam are $\sigma_{zd} = 1.0$ and $\sigma_{zt} = 0.25$, respectively. The beam separation is $d = 4.0$. Fig.~\ref{f1} (b) shows the comparison of the on-axis $E_z$ lineouts from the wake driven by one cell wide beams and beams with $\sigma_r = 0.1$, and they are almost identical.
Therefore, we can ignore the beam spot size and find $\Lambda_t$ for the optimal beam-loading with given $\Lambda_d$, $\sigma_{zd}$, $\sigma_{zt}$ and $d$.
\vspace{0.2cm}

The goal of the optimization is to achieve the minimal energy spread for the trailing beam in the blowout regime, which requires the trailing beam feels the $E_z$ that is as flat as possible in the longitudinal direction. We use the following objective function for the optimization,
\begin{equation}
    F(\Lambda_t) = \sqrt{\frac{\int_{\xi_s}^{\xi_e}(E_z(\xi))^2\lambda_{bt}(\xi)d\xi}{\int_{\xi_s}^{\xi_e}\lambda_{bt}(\xi)d\xi}-\left(\frac{\int_{\xi_s}^{\xi_e} E_z(\xi)\lambda_{bt}(\xi)d\xi}{\int_{\xi_s}^{\xi_e}\lambda_{bt}(\xi)d\xi}\right)^2},\label{eq1}
\end{equation}
where $\xi_s$\ ($\xi_e$) is the head (tail) location of the trailing beam, $\lambda_{bt}(\xi) = \int\rho_{bt}(x, y,\xi)dxdy$ is the normalized charge per unit length of the trailing beam, $\rho_{bt}$ is the normalized charge density of the trailing beam and $\Lambda_t$ is the peak value of $\lambda_{bt}(\xi)$. $F(\Lambda_t) $ is the mean square deviation of weighted on-axis $E_z$, where the density profile of the trailing beam is used as the weight. This is a single-objective optimization~\cite{opti} process because we aim to find the minimum of $F(\Lambda_t)$ while changing $\Lambda_t$. By doing several tests, we find the optimization is a typical convex optimization~\cite{boyd2004convex}, in which for any two points $\Lambda_{t1}$, $\Lambda_{t2}$ in the domain of $\Lambda_t$ and $m\in (0,1)$ we have $F(m\Lambda_{t1}+(1-m)\Lambda_{t2}) \leq mF(\Lambda_{t1})+(1-m)F(\Lambda_{t2})$. For a convex optimization, the local optimum is the global optimum, and the extreme value is the optimal solution~\cite{boyd2004convex}. Thus, the local optimization algorithm can be applied.

To achieve high performance, we optimize the $F(\Lambda_t)$ with the Broyden-Fletcher-Goldfarb-Shanno (BFGS) algorithm ~\cite{1967Quasi}, which has been extensively used to solve nonlinear optimization problems and has been considered to be the most effective of all quasi-Newton methods~\cite{1973On,1974A,1982Local,1989A,2009Global}. We set $\Lambda_t = \Lambda_d$ as the initial solution for the optimization process. By assuming the wakefield does not evolve, the objective function can be evaluated from one-time-step QuickPIC simulation result (i.e. the static wakefield).

\begin{figure}[htb]
    \centering
    \begin{overpic}
        [width=0.6\textwidth,trim=0 0 20 20,clip]{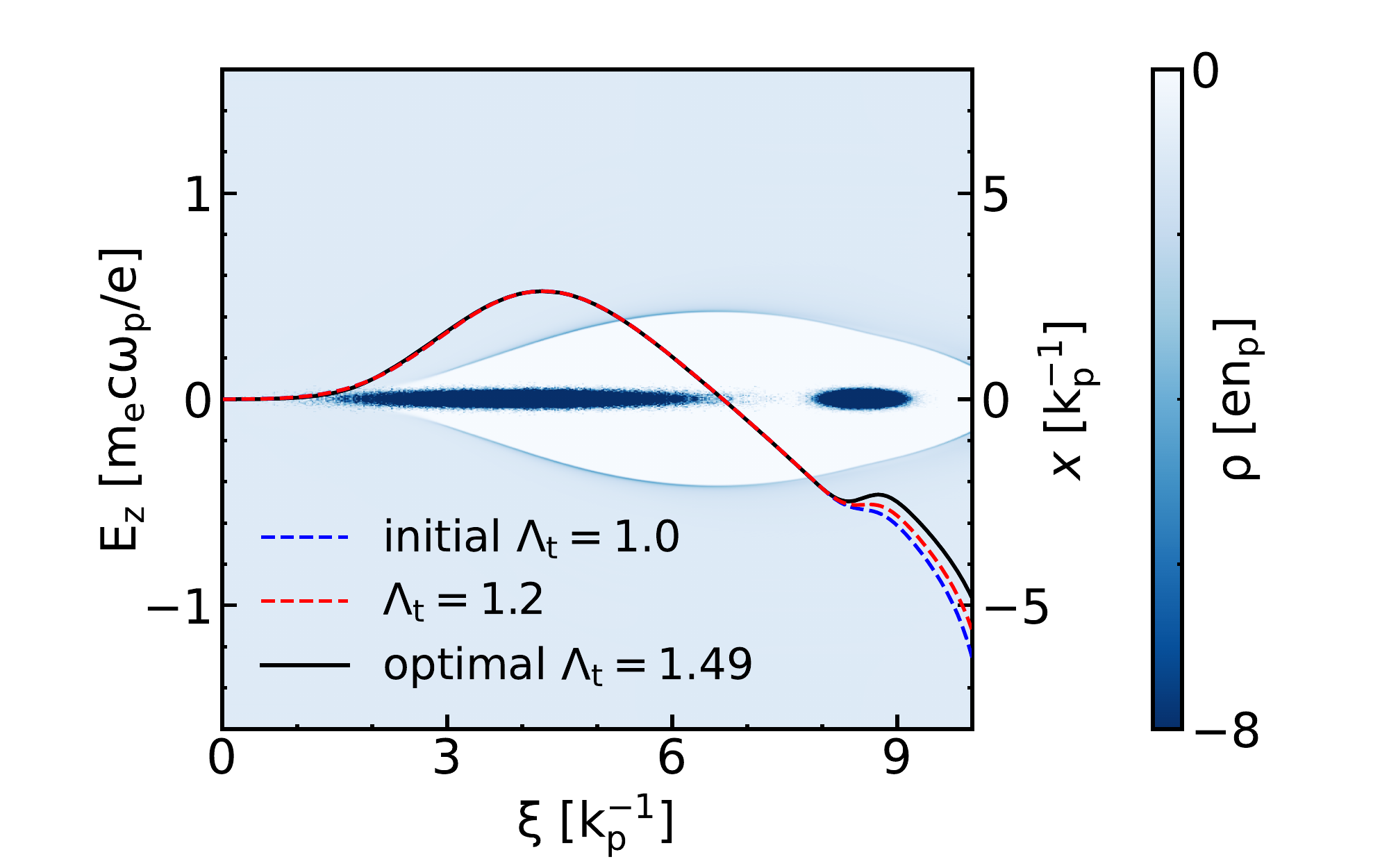}
    \end{overpic}
\renewcommand{\baselinestretch}{1.5}  
\vspace{-0.8cm}
\caption{Accelerating wakefield within the trailing beam before and after an optimization. The blue line represents the initial situation before the optimization and $E_z$ within the trailing beam is underloaded obviously. The black line represents the situation at the end of the optimization and $E_z$ within the trailing beam is flatter than that initially. But the trailing beam has overloaded the wake as seen by the reversal of the $E_z$ slope within the trailing beam. The red line represents the situation that $E_z$ is flat in the middle of the trailing beam. In these simulations, the simulation box is $16.0\times 16.0\times 10.0$ ($x,y,\xi$) and contains $1024\times 1024\times 512$ cells.}\label{f2}
\end{figure}

A typical optimization result is shown in Fig.~\ref{f2}. In this example, beam parameters are $\Lambda_d = 1.0, \sigma_{zd} = 1.0, \sigma_{zt} = 0.25$ and the beam separation is $d = 4.5$. We plot the on-axis $E_z$ at different $\Lambda_t$. The plasma and beam densities are just for illustration, and they do not vary. As shown in Fig.~\ref{f2}, with the optimal $\Lambda_t=1.49$ the trailing beam feels a more flat $E_z$ than that with the initial $\Lambda_t=1.0$ we used. The $E_z$ at the optimal beam-loading is a little overloaded compared with that of $\Lambda_t=1.2$, in which the $\xi$ derivative of $E_z$ only has one zero point within the trailing beam. This is because the trailing beam has a Gaussian profile and the optimal beam-loading case will generate a smaller rms energy spread.
To verify the result obtained from the BFGS algorithm, we manually do a parameter scanning for $\Lambda_t$ from 0.1 to 4.0 with a step size of 0.01. The $\Lambda_t$ for the optimal beam-loading agrees very well with the result from BFGS algorithm. The relative difference between them is about $0.02\%$. With the BFGS algorithm and QuickPIC simulation, the case shown above requires 16 evaluations by QuickPIC to find the optimal $\Lambda_t$, and the total computing time is 7 minutes with 64 cores. We then perform long-distance accelerations. We find that the energy spread of the trailing beam is $1.69\%$ at $\Lambda_t = 1.49$ which is smaller than that with $2.35\%$ at $\Lambda_t = 1.2$ with the same initial energy (10 GeV) and the same energy gain (about 7.3 GeV). This comparison result agrees well with our optimization. We note that it used to be a common sense that the case of red line in Fig.~\ref{f2} would have the smallest rms energy spread. This is not true because the $E_z$ for that case is monotonically decreasing while the black line in Fig.~\ref{f2} is not. As a result, the case of the black line may let more beam particles have the same energy gain at different longitudinal locations, and finally have a smaller rms energy spread than the case of red line.
 
\subsection{Large-range Parameter Scanning for Optimal Beam-loading}
A Python program is developed to automatically optimize a large number of parameter sets of ($\Lambda_d, \sigma_{zd}, \sigma_{zt}, d$) (See Appendix A for details). In these sets of ($\Lambda_d, \sigma_{zd}, \sigma_{zt}, d$), the $\Lambda_d$ has a range of $[0.0144,7.70]$ and the $\sigma_{z}$ for both beams has a range of  $[0.0952,1.90]$. These ranges basically cover the parameters of FACET~\cite{RN15}, FACET II~\cite{RN14}, FLASHForward~\cite{RN136} and other facilities~\cite{RN177} with a plasma density of $10^{16}\ {\rm cm^{-3}}$. When $\Lambda_d$ is set, we scan the $d$ in the range of $[R_{bmax},3R_{bmax}]$, where $R_{bmax}\simeq 2\sqrt{\Lambda_d}$ gives a good estimate of the maximum bubble radius~\cite{RN199}, in order to have the trailing beam be approximately located inside the first plasma bubble. Once the ranges of $\Lambda_d, \sigma_{zd}, \sigma_{zt}$ and $d$ are determined, we evenly 
select values within the range of each parameter. In addition, we also need to ensure that settings for each QuickPIC simulation are appropriate (See Appendix B for details). In each optimization process, we dump the $\Lambda_d, \sigma_{zd}, \sigma_{zt}, d$, the optimal $\Lambda_{t}$, the maximum decelerating wakefield $W_{\rm dec}$ inside the drive beam, the averaged accelerating wakefield felt by the trailing beam $W_{\rm acc} = \int_{\xi_s}^{\xi_e} E_z(\xi)\lambda_{bt}(\xi)d\xi/\int_{\xi_s}^{\xi_e} \lambda_{bt}(\xi)d\xi$ and the transformer ratio $R = \lvert W_{\rm acc}/W_{\rm dec}\rvert$.

Data from the automatic optimization will have some bad parameter sets, i.e. the outliers. For example, some datasets have the trailing beam too far away from the drive beam so that it cannot be effectively accelerated even though the optimization process succeeds. Therefore, we use the boxplot method~\cite{box} and standard normal distribution method~\cite{RN209} to eliminate these outliers. We finally obtain 8537 sets of data for the optimal beam-loading database. The average time for each optimization is only around 7.6 minutes with 64 cores. The range of $\Lambda_d, \sigma_{zd}, \sigma_{zt},d$ and $\Lambda_t$ is presented in Table~\ref{ta1}. Note that Table~\ref{ta1} shows the global range for the beam separation. The actual range of the beam separation varies according to the beam parameters.
\newcolumntype{C}[1]{>{\centering\arraybackslash}p{#1}}
\begin{spacing}{1.5}
\begin{table}[htb]\small
\caption{\label{ta1}
Parameters range obtained from automatic optimizations.}
\renewcommand{\arraystretch}{1.5}
{\begin{tabular}{C{0.35\textwidth}C{0.35\textwidth}} \toprule
Parameters & Range \\ \colrule
$\Lambda_d$ & $[0.0885,7.70]$ \\
$\sigma_{zd}\ [k_p^{-1}]$ & $[0.0952,1.90]$ \\
$d\ [k_p^{-1}]$ & $[1.60,11.1]$ (the global range)\\
$\sigma_{zt}\ [k_p^{-1}]$ & $[0.0952,0.857]$\\
$\Lambda_t$ & $[0.0627,3.14]$ \\ \botrule
\end{tabular}}
\end{table}
\end{spacing}

\section{The Fitting Formulas for Optimal Beam-loading}\label{res}
\subsection{A Data-driven Method}
We use the data-driven method to solve the optimal beam-loading problem in the blowout regime. To obtain explicit fitting formulas, we use the Python library scikit-learn~\cite{RN200} to carry out polynomial regression, which can be generalized into the linear regression~\cite{1972Generalized}. 

During the process, the data are split into several random but with general equal-size folds. And we set some of them as the training dataset and the remaining as the test dataset. Then constructing polynomial features is demanded because the degree of polynomial features we choose directly affects the goodness of fit. Here, we use the coefficient of determination $r^2$~\cite{RN200} to measure how well unseen test dataset tends to be predicted by the model. The closer $r^2$ is to 1, the better the goodness of fit is. To determine the best choice of degree, we use the $k$-fold cross-validation method to evaluate our model to avoid over-fitting~\cite{RN200}. It divides the training dataset into $k$ subsets at once and then trains a model $k$ times in total. In each model training, we use $k-1$ subsets to train the model and use the remaining one to validate the model and obtain the $r^2$ for each training. The averaged $r^2$ is obtained at the end of this loop for a particular degree. And the best degree should have the largest averaged $r^2$ with this $k$-fold cross-validation method. As a common choice, we choose $k=10$ for our calculation. After determining the best degree, we use the whole training dataset to train a model (i.e. get the fitting formula) and use the test dataset to do the final evaluation.  

\subsection{The Fitting Formulas for $\Lambda_t$ and $R$}
By using the method described above, we can obtain the fitting formula for the optimal $\Lambda_t$, which can be written as $\Lambda_t = f(\Lambda_d,\sigma_{zd},\sigma_{zt},d)$. More specifically, training dataset and test dataset account for 75\% and 25\% of the database, respectively. When we use training dataset to perform 10-fold cross-validation, we obtain the averaged $r^2\simeq0.999$ at degree of 3, which is larger than that at other degree values. Therefore, we use the whole training dataset to do the polynomial regression at degree of 3 and obtain $r^2\simeq0.999$ when evaluating  the test dataset. This represents high prediction accuracy. The final result of the polynomial regression, i.e. the fitting formula for $\Lambda_t$, is
\begin{widetext}
\begin{align}
\begin{split}
    \Lambda_t &= h_0+h_1\Lambda_d+h_2\sigma_{zd}+h_3\sigma_{zt}+h_4d+h_5\Lambda_d^2+h_6\Lambda_d\sigma_{zd}+h_7\Lambda_d\sigma_{zt}+h_8\Lambda_dd\\
    &+h_9\sigma_{zd}^2++h_{10}\sigma_{zd}\sigma_{zt}+h_{11}\sigma_{zd}d+h_{12}\sigma_{zt}^2+h_{13}\sigma_{zt}d+h_{14}d^2+h_{15}\Lambda_d^3+h_{16}\Lambda_d^2\sigma_{zd}\\
    &+h_{17}\Lambda_d^2\sigma_{zt}+h_{18}\Lambda_d^2d+h_{19}\Lambda_d\sigma_{zd}^2+h_{20}\Lambda_d\sigma_{zd}\sigma_{zt}+h_{21}\Lambda_d\sigma_{zd}d+h_{22}\Lambda_d\sigma_{zt}^2\\
    &+h_{23}\Lambda_d\sigma_{zt}d+h_{24}\Lambda_dd^2+h_{25}\sigma_{zd}^3+h_{26}\sigma_{zd}^2\sigma_{zt}+h_{27}\sigma_{zd}^2d+h_{28}\sigma_{zd}\sigma_{zt}^2\\
    &+h_{29}\sigma_{zd}\sigma_{zt}d+h_{30}\sigma_{zd}d^2+h_{31}\sigma_{zt}^3+h_{32}\sigma_{zt}^2d+h_{33}\sigma_{zt} d^2+h_{34}d^3,\label{eq2}
\end{split}
\end{align}
\end{widetext}
where the fitting coefficients are given in Table~\ref{ta2}.
\begin{table}[htbp]\small
\caption{\label{ta2}
Fitting coefficients for the fitting formula of $\Lambda_t$.}
\renewcommand{\arraystretch}{1.5}
\setlength{\tabcolsep}{0.3mm}{
\begin{ruledtabular}
\begin{tabular}{lllllllll}
$h_0$=-5.014$\times 10^{-1}$ & $h_1$=3.658$\times 10^{-1}$ & $h_2$=9.119$\times 10^{-1}$ & $h_3$=-1.083 & $h_4$=3.062$\times 10^{-1}$\\
$h_5$=-3.754$\times 10^{-2}$ & $h_6$=2.344 & $h_7$=1.281$\times 10^{-1}$ & $h_8$=-5.028$\times 10^{-2}$ & $h_9$=-7.136$\times 10^{-1}$\\
$h_{10}$=-1.915$\times 10^{-1}$ & $h_{11}$=-1.316$\times 10^{-1}$ & $h_{12}$=-2.167 & $h_{13}$=1.034 & $h_{14}$=-7.607$\times 10^{-2}$\\
$h_{15}$=-2.391$\times 10^{-3}$ & $h_{16}$=-7.570$\times 10^{-2}$ & $h_{17}$=2.641$\times 10^{-2}$ & $h_{18}$=1.160$\times 10^{-2}$ & $h_{19}$=-8.626$\times 10^{-1}$\\ 
$h_{20}$=-2.424$\times 10^{-1}$ & $h_{21}$=9.630$\times 10^{-2}$ & $h_{22}$=3.874$\times 10^{-1}$ & $h_{23}$=-7.137$\times 10^{-2}$ & $h_{24}$=-3.061$\times 10^{-3}$\\ 
$h_{25}$=1.238$\times 10^{-1}$ & $h_{26}$=3.752$\times 10^{-2}$ & $h_{27}$=7.655$\times 10^{-2}$ & $h_{28}$=5.197$\times 10^{-1}$ & $h_{29}$=-2.585$\times 10^{-2}$\\
$h_{30}$=-6.071$\times 10^{-3}$ & $h_{31}$=-2.866 & $h_{32}$=1.231 & $h_{33}$=-2.525$\times 10^{-1}$ & $h_{34}$=6.674$\times 10^{-3}$\\
\end{tabular}
\end{ruledtabular}}
\end{table}

Besides $\Lambda_t$, the transformer ratio $R$ is also an important parameter we concern in a two-bunch PWFA stage. We consider that $R$ is dependent on $\Lambda_d, \Lambda_t, \sigma_{zd}, \sigma_{zt}$ and $d$. Following the same procedure, we can get the explicit expression of $R = f(\Lambda_d,\Lambda_t,\sigma_{zd},\sigma_{zt},d)$. In this case, training dataset and test dataset comprise 80\% and 20\% of the whole database, respectively. We finally choose the degree of 2, with which we get the highest averaged $r^2\simeq0.98$ when performing $10$-fold cross-validation. In the final evaluation using the test dataset, we get $r^2\simeq0.99$, which represents high prediction accuracy. The fitting formula for $R$ is
\begin{widetext}
\begin{align}
\begin{split}
    R &= p_0+p_1\Lambda_d+p_2\sigma_{zd}+p_3\sigma_{zt}+p_4d+p_5\Lambda_t+p_6\Lambda_d^2+p_7\Lambda_d\sigma_{zd}+p_8\Lambda_d\sigma_{zt}\\
    &+p_9\Lambda_dd+p_{10}\Lambda_d\Lambda_t+p_{11}\sigma_{zd}^2+p_{12}\sigma_{zd}\sigma_{zt}+p_{13}\sigma_{zd}d+p_{14}\sigma_{zd}\Lambda_t\\
    &+p_{15}\sigma_{zt}^2+p_{16}\sigma_{zt}d+p_{17}\sigma_{zt}\Lambda_t+p_{18}d^2+p_{19}d\Lambda_t+p_{20}\Lambda_t^2,\label{eq3}
\end{split}
\end{align}
\end{widetext}
where the fitting coefficients are given in Table~\ref{ta3}.
\begin{table}[htbp]\small
\caption{\label{ta3}
Fitting coefficients for the fitting formula of $R$.}
\renewcommand{\arraystretch}{1.5}
\begin{ruledtabular}
\begin{tabular}{lllllll}
$p_0$=-1.453 & $p_1$=0.3199 & $p_2$=0.3178 & $p_3$=0.3084 & $p_4$=0.7241 & $p_5$=-0.8454 & $p_6$=0.02719\\
$p_7$=0.4858 & $p_8$=0.4140 & $p_9$=-0.1070 & $p_{10}$=-0.02761 & $p_{11}$=-0.2779 & $p_{12}$=-0.4929 & $p_{13}$=0.2440\\
$p_{14}$=-0.3681 & $p_{15}$=1.632 & $p_{16}$=-0.6407 & $p_{17}$=-0.01004 & $p_{18}$=-0.01716 & $p_{19}$=0.01431 & $p_{20}$=0.1439\\
\end{tabular}
\end{ruledtabular}
\end{table}

\begin{figure}[htb]
    \centering
    \begin{overpic}
        [width=0.49\textwidth,trim=0 0 20 20,clip]{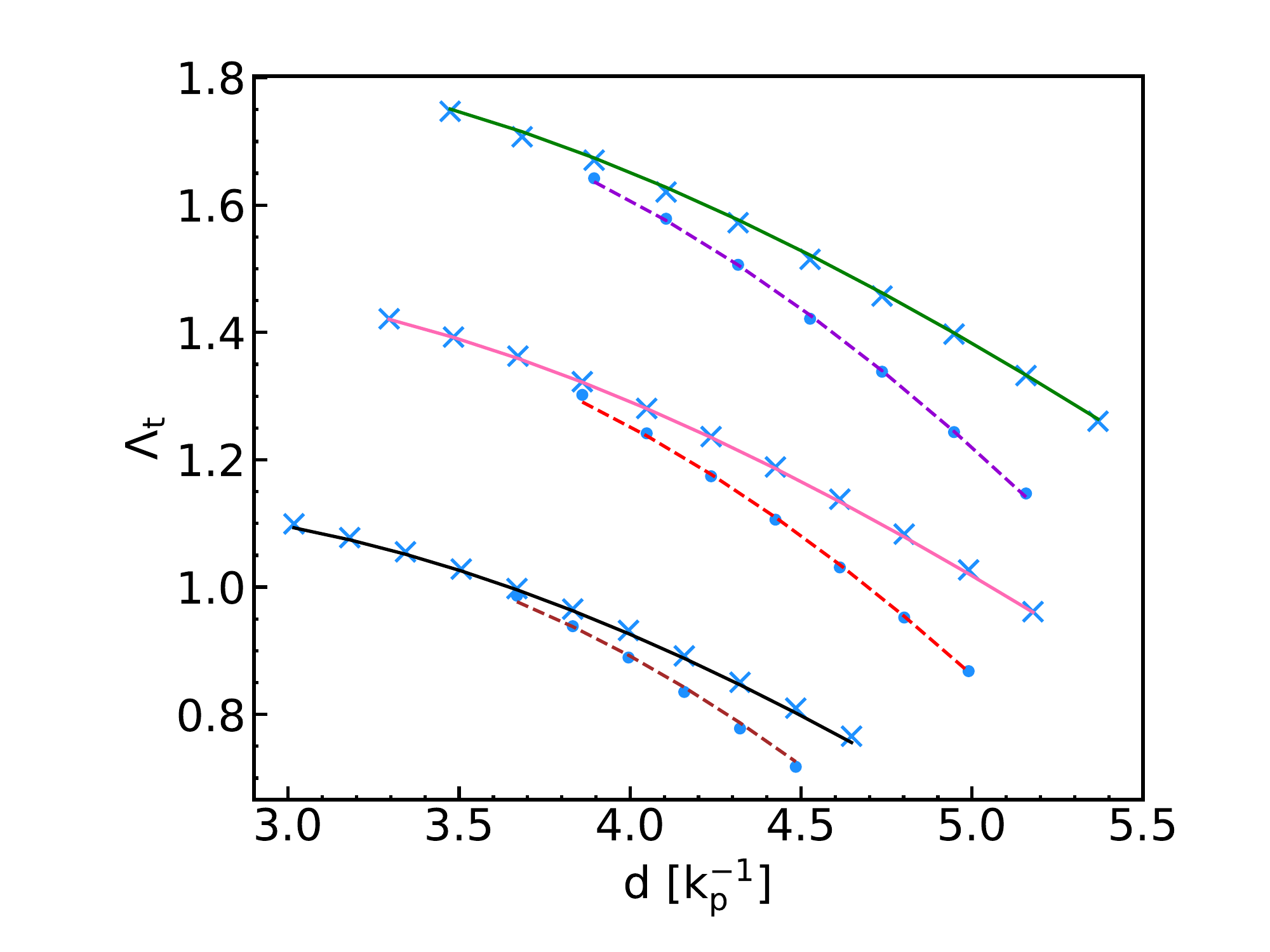}
        \put(22,65){\small{(a)}}
    \end{overpic}
    \begin{overpic}
        [width=0.46\textwidth]{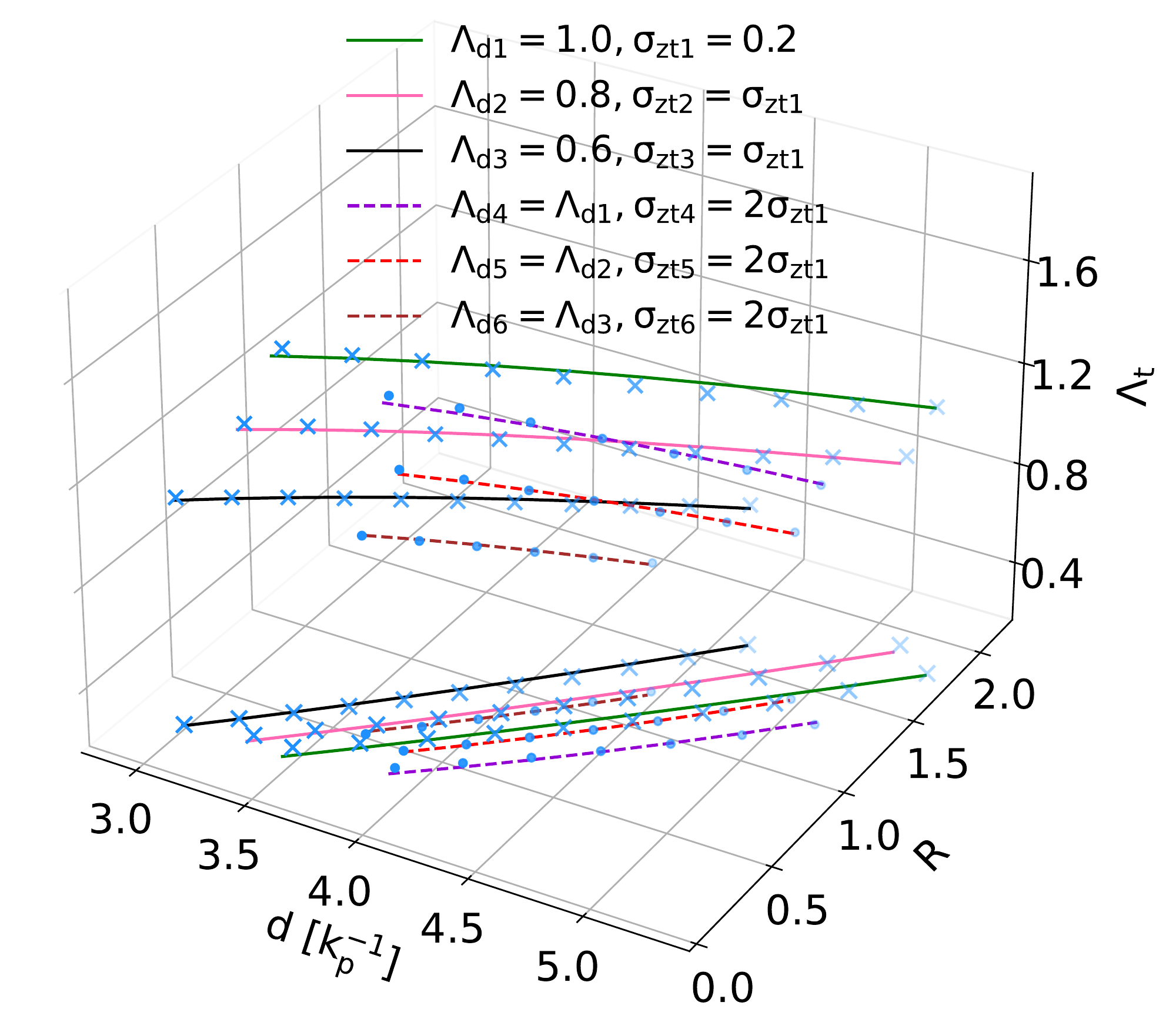}
        \put(8,71){\small{(b)}}
    \end{overpic}
\renewcommand{\baselinestretch}{1.5}
 \caption{(a) The Optimal $\Lambda_t$ versus the beam separation $d$. The blue cross and dot points are the results from the optimization program. The solid and dashed lines are the results from Eq.~(\ref{eq2}). All the solid and dashed lines have the same $\sigma_{zd}=1.0$ but different $\Lambda_d$ or $\sigma_{zt}$. (b) The Optimal $\Lambda_t$ versus the beam separation $d$ and the transformer ratio $R$, and the projected lines on the $d - R$ plane. Each line has the same $\Lambda_d$, $\sigma_{zd}$ and $\sigma_{zt}$ as those in (a).}\label{f3}
\end{figure}

Through the fitting formulas, we can obtain the optimal $\Lambda_{t}$ without running the optimization program. For example, for $\Lambda_d = 1.0, \sigma_{zd} = 1.0, \sigma_{zt} = 0.2$ and $d = 4.0$, Eq.~(\ref{eq2}) gives the optimal $\Lambda_{t} = 1.652$, while the optimization program gives $\Lambda_{t} = 1.644$. The results agree well with each other. When calculating the transformer ratio $R$ using the fitting formula, we first need to obtain the optimal $\Lambda_{t}$ through Eq.~(\ref{eq2}), and then substitute the optimal $\Lambda_{t}$ into Eq.~(\ref{eq3}) to obtain $R$. This gives $R=0.622$ in this case, while the optimization program gives $R=0.622$. They still agree very well with each other. In Fig.~\ref{f3}, we compare more results from the optimization program with the results given by the fitting formulas.
The green solid line in Fig.~\ref{f3} (a) plots the optimal $\Lambda_t$ versus $d$ with $\Lambda_d = 1.0, \sigma_{zd} = 1.0$ and $\sigma_{zt} = 0.2$ by using the fitting formula Eq.~(\ref{eq2}). The blue cross points are the results from the optimization program and they agree very well with the fitting results. The pink and black solid lines in Fig.~\ref{f3} (a) have different $\Lambda_{d}$ but the same $\sigma_{zd}$ and $\sigma_{zt}$, and they agree very well with the results from the optimization program. We also change $\sigma_{zt}$ and $\Lambda_{d}$ while keeping $\sigma_{zd}$ and still find good agreements between the fitting results (dashed lines) and the optimization results (dot points) as shown in Fig.~\ref{f3} (a). Furthermore, we calculate the transformer ratio $R$ from the fitting formula Eq.~(\ref{eq3}) and show the results in Fig.~\ref{f3} (b), which has another axis of $R$ than Fig.~\ref{f3} (a). The fitting results also agree very well with the optimization results. From the results shown in Fig.~\ref{f3}, we can also find that for given $\Lambda_d, \sigma_{zd}$ and $\sigma_{zt}$, the bigger the $d$ is, the smaller the $\Lambda_t$ is and the higher the $R$ is, which agrees with the understanding of beam-loading in the nonlinear plasma wake~\cite{RN152}. The applicable parameter range for these two fitting formulas is listed in Table~\ref{ta1}. In addition, the beam energy had better to be larger than $100\ \rm{MeV}$ when using these fitting formulas.

\subsection{Flat-top and Trapezoidal Trailing Beams}\label{extra}
\begin{figure}[htb]
    \centering
    \begin{overpic}
        [width=0.55\textwidth,trim=15 0 30 0,clip]{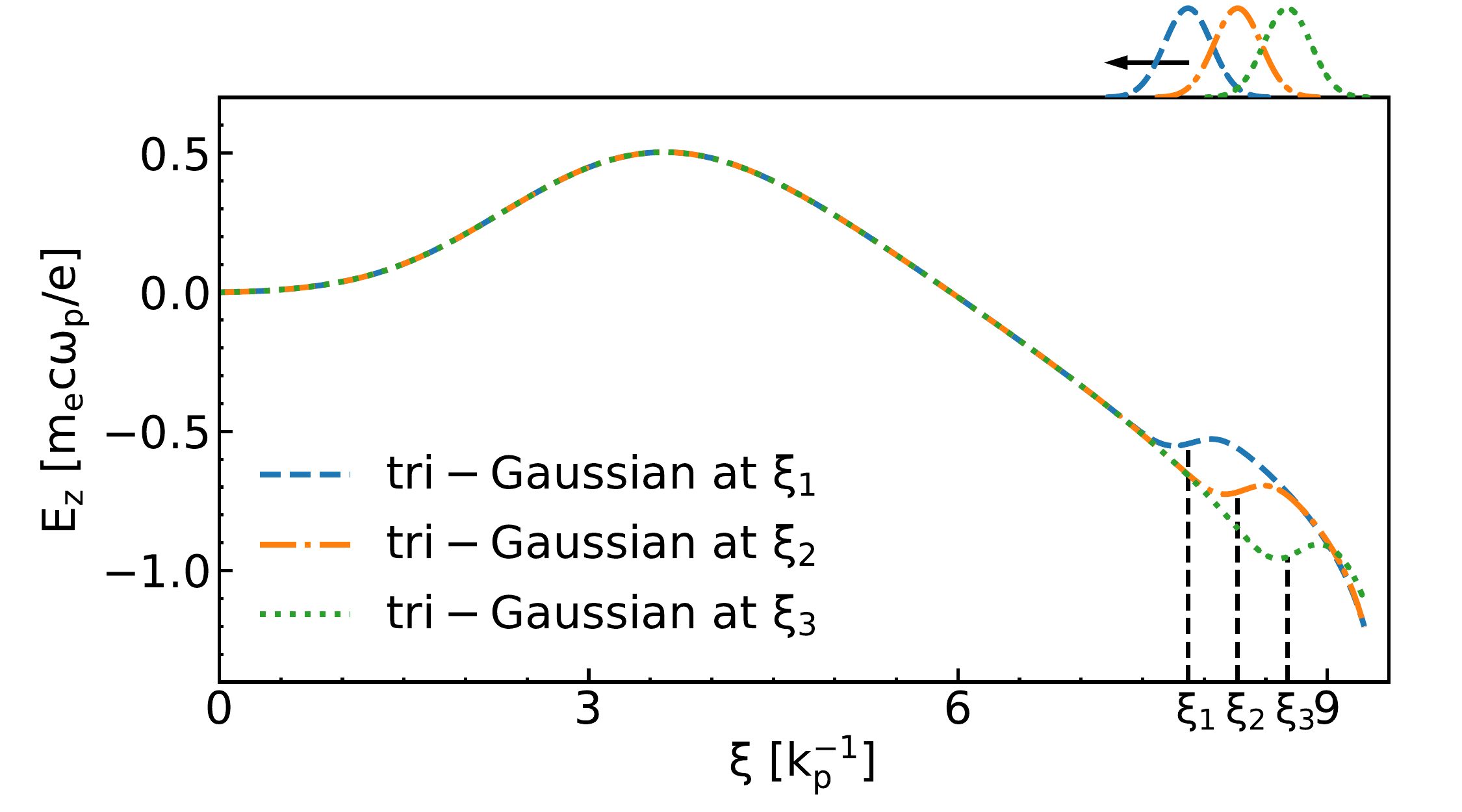}
        \put(15,48){\small{(a)}}
    \end{overpic}
    \begin{overpic}
        [width=0.55\textwidth,trim=15 0 30 0,clip]{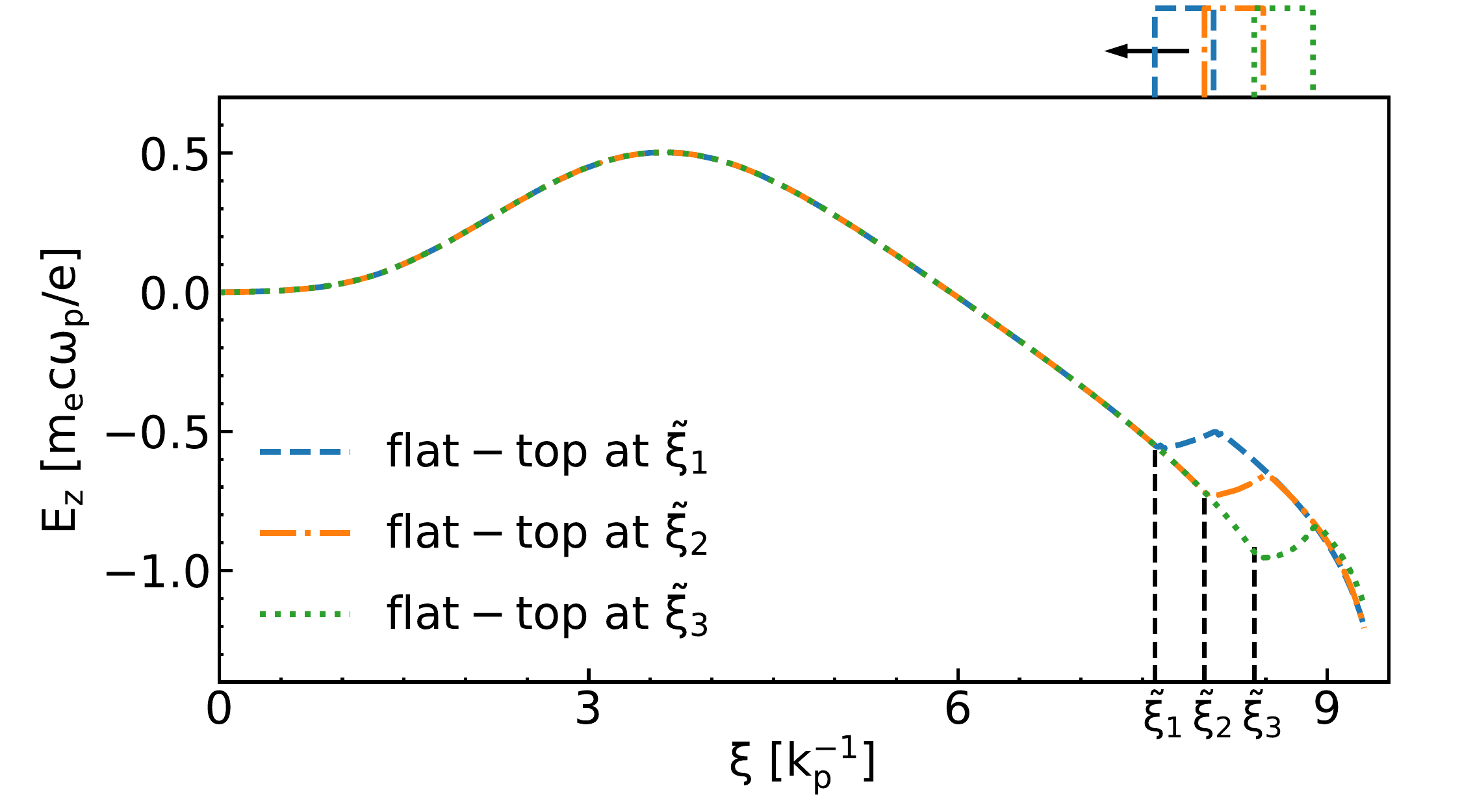}
        \put(15,48){\small{(b)}}
    \end{overpic}
    \begin{overpic}
        [width=0.55\textwidth,trim=15 0 30 0,clip]{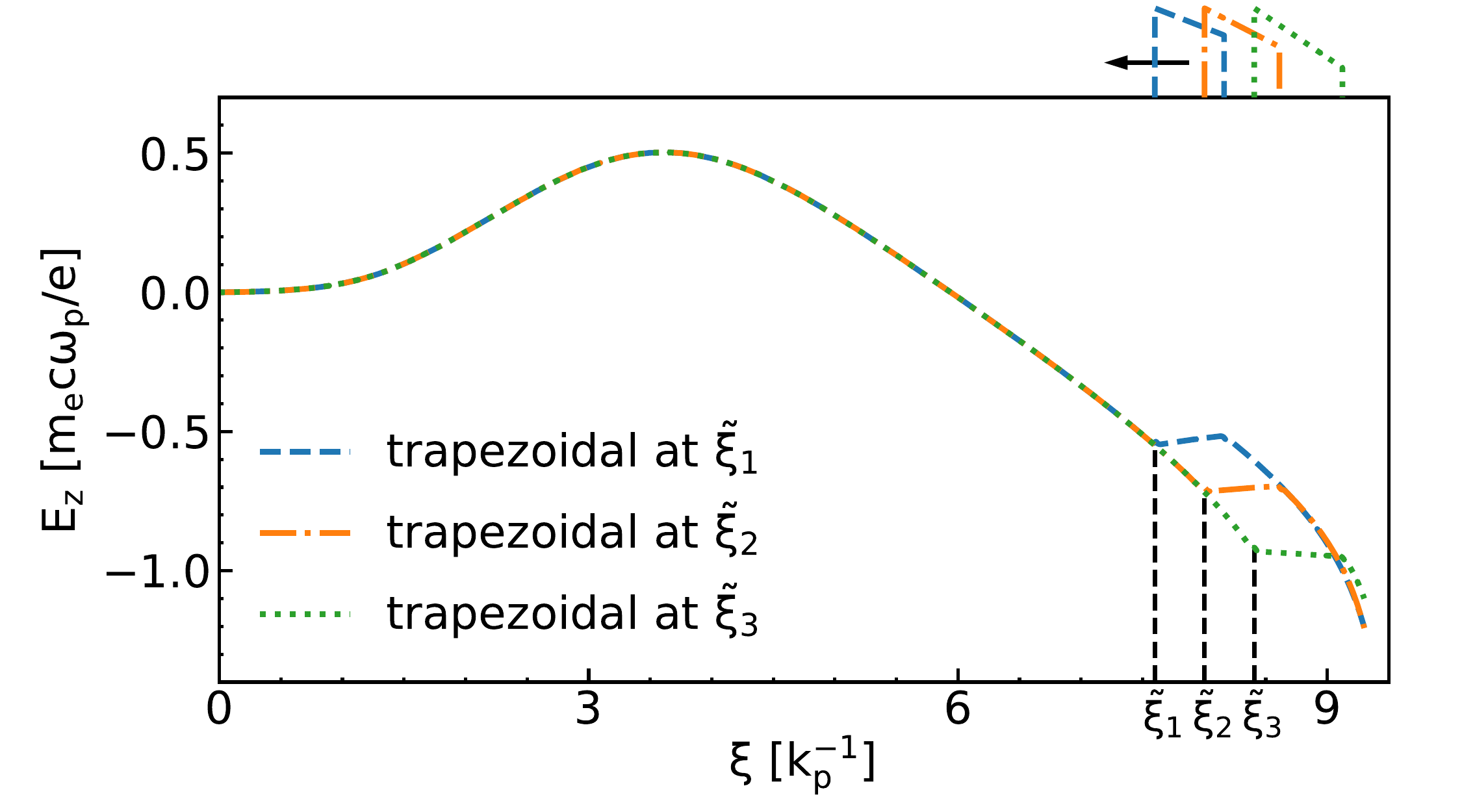}
        \put(15,48){\small{(c)}}
    \end{overpic}
\renewcommand{\baselinestretch}{1.5}
 \caption{The lineout of longitudinal wakefield $E_z$ for (a) tri-Gaussian beams, (b) flat-top beams and (c) trapezoidal beams at three distinct locations.}\label{f4}
\end{figure}

We also test the availability of the fitting formulas for trailing beams with a longitudinal flat-top profile or a longitudinal trapezoidal profile. We pick up three tri-Gaussian cases with the same drive beam parameters and the same $\sigma_{zt} = 0.190$ but different $d$. We plot the on-axis $E_z$ of the plasma wake in Fig.~\ref{f4} (a). 
In these simulations, the drive beam has $\Lambda_d = 0.918$ and $\sigma_{zd} = 0.952$, and its beam center is located at $\xi_d = 3.33$. For each simulation as shown in Fig.~\ref{f4} (a), a tri-Gaussian trailing beam is loaded at three distinct locations, $\xi_1=\xi_d+4.538, \xi_2=\xi_d+4.942$ and $\xi_3=\xi_d+5.345$. According to Eq.~(\ref{eq2}), three optimal $\Lambda_t$ are $\Lambda_{t1}=1.345$ at $d=4.538$, $\Lambda_{t2}=1.229$ at $d=4.942$ and $\Lambda_{t3}=1.104$ at $d=5.345$. As shown in Fig.~\ref{f4} (a), all these three cases have reached the optimal beam-loading. When switching them to the longitudinal flat-top profile, we keep $\Lambda_t$ and the total particle number the same as those of tri-Gaussian trailing beams. Therefore, the flat-top beam length should be $l_{zF} = \sqrt{2\pi}\sigma_{zt}$. We load these flat-top beams with their heads at a distance $\sqrt{2}\sigma_{zt}$ in front of $\xi_{1,2,3}$ in order to maintain the transformer ratio (as suggested in Ref.~[\onlinecite{RN152}]). As shown in Fig.~\ref{f4} (b), the beam-loading effect of flat-top trailing beams mimics that of tri-Gaussian trailing beams. In Ref.~[\onlinecite{RN152}], it is shown that the trapezoidal trailing beams can perfectly flatten the $E_z$. For trapezoidal trailing beams, we still keep the total particle number and maximal $\Lambda_t$ the same as those of tri-Gaussian trailing beams. The trapezoidal beam also has a sharp edge as the flat-top beam. Thus, we load trapezoidal beams at $\tilde{\xi}_{1,2,3}=\xi_{1,2,3}-\sqrt{2}\sigma_{zt}$. The slope of the trapezoidal profile $a$ equals to $E_z$ where the beam-loading starts~\cite{RN152}, which roughly equals to the averaged accelerating wakefield of the tri-Gaussian beam. For three trapezoidal trailing beams plotted in Fig.~\ref{f4} (c), we have $a_{1}=-0.539$, beam length $l_{z1}=0.562$ at $\tilde{\xi}_1$, $a_{2}=-0.709,\ l_{z2}=0.609$ at $\tilde{\xi}_2$ and $a_{3}=-0.932,\ l_{z3}=0.716$ at $\tilde{\xi}_3$, where the beam length is derived from the total charge of the beam. As shown in Fig.~\ref{f4} (c), $E_z$ is almost flattened and the transformer ratio is well maintained. Therefore, through proper beam parameter transformations, the fitting formulas of tri-Gaussian beams can still give a good estimation for flat-top or trapezoidal trailing beams.

\section{A Scaling Law for Charges of Two Beams Under the Optimal Beam-loading}\label{charge}
Not only can the fitting formulas be used to find particular beam parameters for the optimal beam-loading, they can also unveil many physics features under the optimal beam-loading. One of the features is the relation between the charge of the drive beam and that of the trailing beam under the optimal beam-loading. The beam charge is proportional to $\Lambda\sigma_z$. Therefore, by multiplying $\sigma_{zt}$ on both sides of Eq.~(\ref{eq2}) and rearranging the right hand side of the equation, we can find the relation between $\Lambda_t\sigma_{zt}$ and $\Lambda_d\sigma_{zd}$ as
\begin{equation}
    \Lambda_t\sigma_{zt} = A\cdot(\Lambda_d\sigma_{zd})^3+B\cdot(\Lambda_d\sigma_{zd})^2+D\cdot(\Lambda_d\sigma_{zd})+G,\label{eq4}
\end{equation}
where $A = h_{15}(\sigma_{zt}/\sigma_{zd}^3),\ 
B = (h_{5}+h_{16}\sigma_{zd}+h_{17}\sigma_{zt}+h_{18}d)(\sigma_{zt}/\sigma_{zd}^2),\ 
D = (h_1+h_6\sigma_{zd}+h_7\sigma_{zt}+h_8d+h_{19}\sigma_{zd}^2+h_{20}\sigma_{zd}\sigma_{zt}+h_{21}\sigma_{zd}d+h_{22}\sigma_{zt}^2+h_{23}\sigma_{zt}d+h_{24}d^2)(\sigma_{zt}/\sigma_{zd}),\ 
G = (h_2\sigma_{zd}+h_3\sigma_{zt}+h_4d+h_9\sigma_{zd}^2+h_{10}\sigma_{zd}\sigma_{zt}+h_{11}\sigma_{zd}d+h_{12}\sigma_{zt}^2+h_{13}\sigma_{zt}d+h_{14}d^2+h_{25}\sigma_{zd}^3+h_{26}\sigma_{zd}^2\sigma_{zt}+h_{27}\sigma_{zd}^2d+h_{28}\sigma_{zd}\sigma_{zt}^2+h_{29}\sigma_{zd}\sigma_{zt}d+h_{30}\sigma_{zd}d^2+h_{31}\sigma_{zt}^3+h_{32}\sigma_{zt}^2d+h_{33}\sigma_{zt}d^2+h_{34}d^3+h_0)\sigma_{zt}$.

\begin{figure}[htb]
    \centering
    \begin{overpic}
        [width=0.49\textwidth,trim=40 0 20 20,clip]{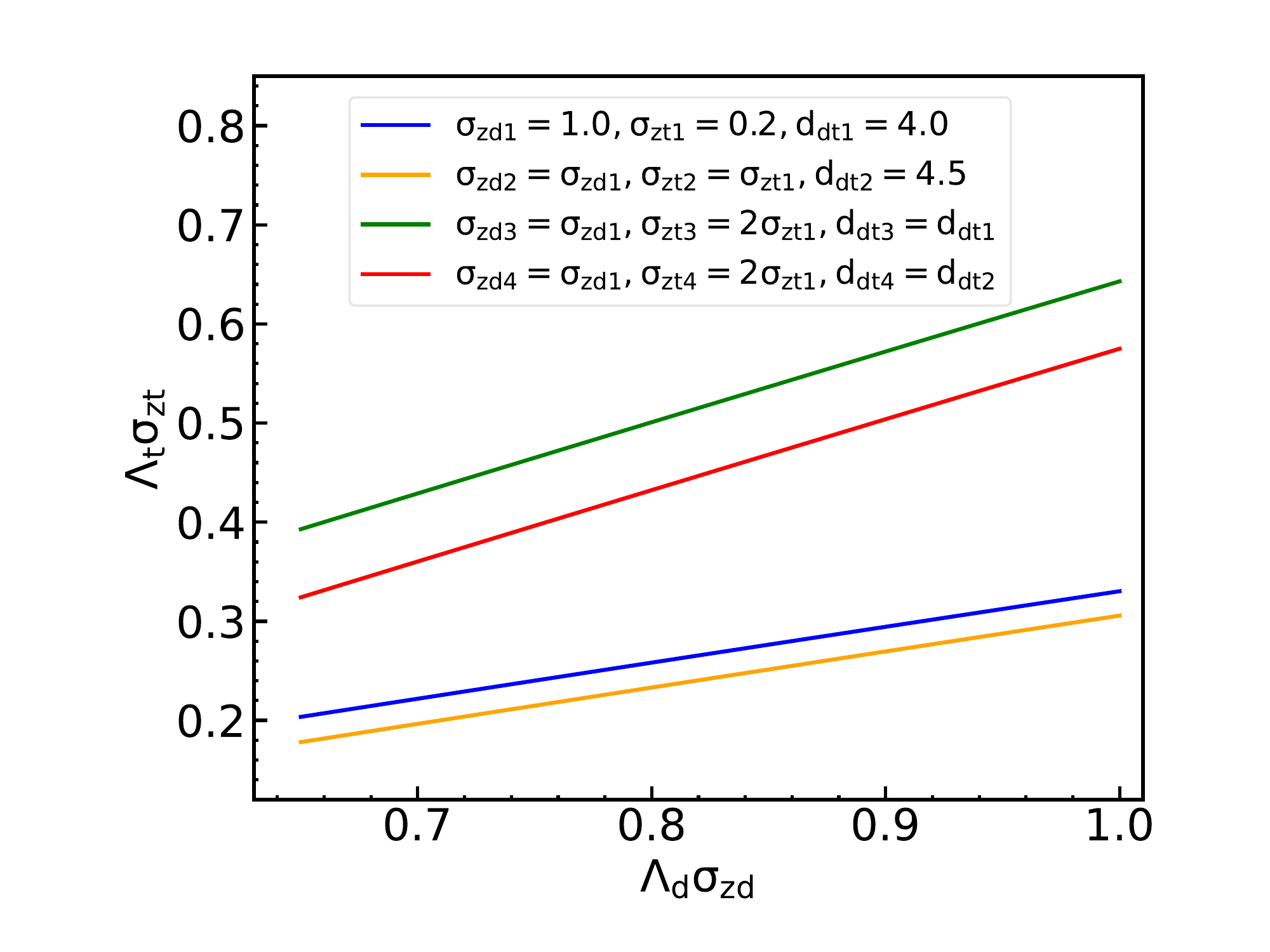}
        \put(16,70){\small{(a)}}
    \end{overpic}
    \begin{overpic}
        [width=0.46\textwidth,trim=10 10 10 10,clip]{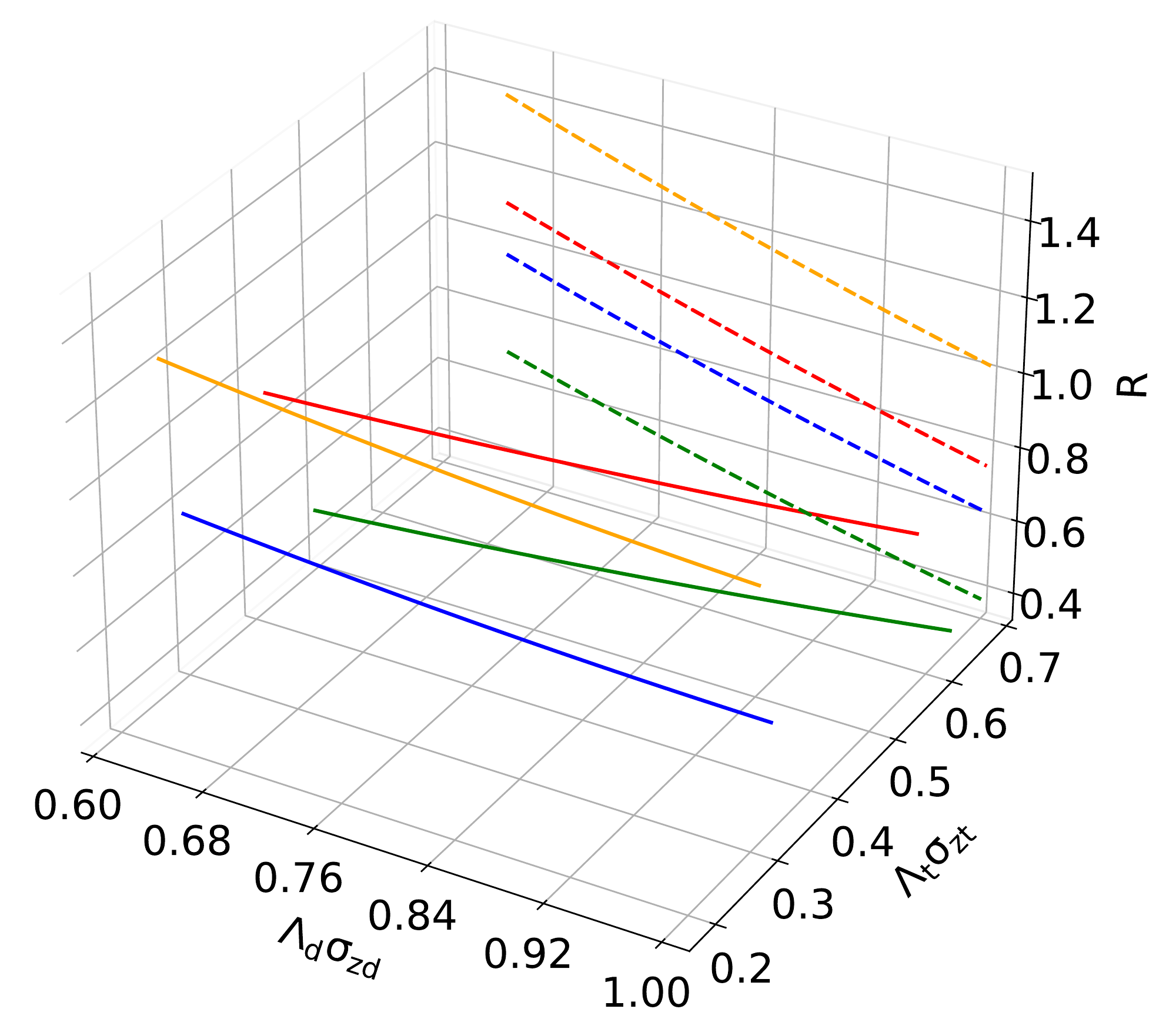}
        \put(5,75){\small{(b)}}
    \end{overpic}
\renewcommand{\baselinestretch}{1.5}
\caption{(a) The relation between $\Lambda_t\sigma_{zt}$ and $\Lambda_d\sigma_{zd}$ under the optimal beam-loading. (b) The transformer ratio $R$ versus $\Lambda_t\sigma_{zt}$ and $\Lambda_d\sigma_{zd}$, and the projected lines on the $\Lambda_d\sigma_{zd} - R$ plane. Each line has the same $\sigma_{zd}, \sigma_{zt}$ and $d$ as those in (a).}\label{f5}
\end{figure}

According to Eq.~(\ref{eq4}), with $\sigma_{zd} = 1.0, \sigma_{zt} = 0.2$ and $d = 4.0$ as an example, we can get $\Lambda_t\sigma_{zt} = -0.0004781(\Lambda_d\sigma_{zd})^3-0.01231(\Lambda_d\sigma_{zd})^2+0.3835(\Lambda_d\sigma_{zd})-0.04041$, which is plotted as the blue line in Fig.~\ref{f5} (a). 
From the plot, we can find that $\Lambda_t\sigma_{zt}$ almost increases linearly with $\Lambda_d\sigma_{zd}$. This is because the high order terms are much less than the $\Lambda_d\sigma_{zd}$ term in this example. Therefore, Eq.~\ref{eq4} can be reduced to $\Lambda_t\sigma_{zt} = D(\Lambda_d\sigma_{zd})+G$. This means that once the optimal beam-loading is reached, it is always satisfied when increasing charges of both beams with the same ratio $D$. In Fig.~\ref{f5} (a), we plot three other lines with different $d$ or $\sigma_{zt}$. And they all obey the simple scaling law $\Lambda_t\sigma_{zt} = D(\Lambda_d\sigma_{zd})+G$, where $D$ and $G$ depend on $\sigma_{zd}, \sigma_{zt}$ and $d$. If G is much less than $D \Lambda_d\sigma_{zd}$, we can further neglect G and $\Lambda_t\sigma_{zt}$ will become proportional to $\Lambda_d\sigma_{zd}$. This means that we can change charges of both beams at the same rate without breaking the optimal beam-loading condition. In addition, with Eq.~(\ref{eq3}) we can also calculate the transformer ratio $R$ for the lines in Fig.~\ref{f5} (a), which is shown in Fig.~\ref{f5} (b). This will bring much convenience for designing a two-bunch PWFA stage.

\section{Optimal Plasma Densities for Maximum Accelerated Charge and Maximum Acceleration Efficiency}\label{eng}
So far, we are using the normalized units for each parameter. This means the physics features we obtained in the last section is only available for a fixed plasma density. However, we are also interested in how the beam parameter varies as the plasma density changes under the optimal beam-loading. This can be obtained by switching the normalized units in the equation back to the original ones. We have the charge of the drive (trailing) beam $Q_d\ (Q_t)[nC] = (2\pi)^{\frac{3}{2}}en_pk_p^{-3}\Lambda_{d(t)} \sigma_{zd(t)} = 3.79/\sqrt{n_p[10^{16}cm^{-3}]}\Lambda_{d(t)} \sigma_{zd(t)}$, the rms length of the drive (trailing) beam $L_d\ (L_t)[\mu m] = \sigma_{zd(t)}k_p^{-1}= 53.14/\sqrt{n_p[10^{16}cm^{-3}]}\sigma_{zd(t)}$ and the beam separation $l[\mu m] = d\cdot k_p^{-1} = 53.14/\sqrt{n_p[10^{16}cm^{-3}]}d$. Therefore, Eq.~(\ref{eq2}) and Eq.~(\ref{eq3}) can be converted to equations that have the plasma density as an additional variable (See Appendix C and Eq.~(\ref{eq6}) for details). Here we will focus on how the plasma density will affect Eq.~(\ref{eq4}). We convert Eq.~(\ref{eq4}) into an engineering formula
\begin{equation}
    Q_t[{\rm nC}] = H\cdot n_p^{\frac{3}{2}}[{\rm 10^{16} cm^{-3}}]+M\cdot n_p[{\rm 10^{16} cm^{-3}}]+P\cdot n_p^{\frac{1}{2}}[{\rm 10^{16}cm^{-3}}]+S,\label{eq5}
\end{equation}
where $H = (w_{25}L_d^3+w_{26}L_d^2 L_t+w_{27}L_d^2 l+w_{28}L_d L_t^2+w_{29}L_d L_t l+w_{30}L_d l^2+w_{31}L_t^3+w_{32}L_t^2 l+w_{33}L_t l^2+w_{34}l^3)L_t,\ 
M = (w_9L_d^2+w_{10}L_d L_t+w_{11}L_d l+w_{12}L_t^2+w_{13}L_t l+w_{14}l^2)L_t+
(w_{19}L_d^2+w_{20}L_d L_t+w_{21}L_d l+w_{22}L_t^2+w_{23}L_t l+w_{24}l^2)L_t Q_d/L_d,\
P = (w_2L_d+w_3L_t+w_4l)L_t+(w_6L_d+w_7L_t+w_8l)L_t Q_d/L_d+(w_{16}L_d+w_{17}L_t+w_{18}l)L_t Q_d^2/L_d^2$ and $S = w_1L_t Q_d/L_d+w_5L_t Q_d^2/L_d^2+w_{15}L_t Q_d^3/L_d^3+w_0L_t$. The coefficients in Eq.~(\ref{eq5}) are listed in Table~\ref{ta4}.

\begin{table}[htbp]\small
\caption{\label{ta4}
Coefficients in Eq.~(\ref{eq5}).}
\renewcommand{\arraystretch}{1.5}
\setlength{\tabcolsep}{0.3mm}{
\begin{ruledtabular}
\begin{tabular}{lllllllll}
$w_0$=-3.573$\times 10^{-2}$ & $w_1$=3.658$\times 10^{-1}$ & $w_2$=1.223$\times 10^{-3}$ & $w_3$=-1.452$\times 10^{-3}$ & $w_4$=4.107$\times 10^{-4}$ \\
$w_5$=-5.268$\times 10^{-1}$ & $w_6$=4.411$\times 10^{-2}$ & $w_7$=2.410$\times 10^{-3}$ & $w_8$=-9.462$\times 10^{-4}$ & $w_9$=-1.801$\times 10^{-5}$\\
$w_{10}$=-4.832$\times 10^{-6}$ & $w_{11}$=-3.321$\times 10^{-6}$ & $w_{12}$=-5.468$\times 10^{-5}$ & $w_{13}$=2.610$\times 10^{-5}$ & $w_{14}$=-1.919$\times 10^{-6}$\\ 
$w_{15}$=-4.708$\times 10^{-1}$ & $w_{16}$=-1.999$\times 10^{-2}$ & $w_{17}$=6.975$\times 10^{-3}$ & $w_{18}$=3.062$\times 10^{-3}$ & $w_{19}$=-3.055$\times 10^{-4}$\\ 
$w_{20}$=-8.582$\times 10^{-5}$ & $w_{21}$=3.410$\times 10^{-5}$ & $w_{22}$=1.372$\times 10^{-4}$ & $w_{23}$=-2.527$\times 10^{-5}$ & $w_{24}$=-1.084$\times 10^{-6}$\\
$w_{25}$=5.880$\times 10^{-8}$ & $w_{26}$=1.781$\times 10^{-8}$ & $w_{27}$=3.635$\times 10^{-8}$ & $w_{28}$=2.468$\times 10^{-7}$ & $w_{29}$=-1.227$\times 10^{-8}$\\ 
$w_{30}$=-2.883$\times 10^{-9}$ & $w_{31}$=-1.361$\times 10^{-6}$ & $w_{32}$=5.844$\times 10^{-7}$ & $w_{33}$=-1.199$\times 10^{-7}$ & $w_{34}$=3.169$\times 10^{-9}$ & \\
\end{tabular}
\end{ruledtabular}}
\end{table}

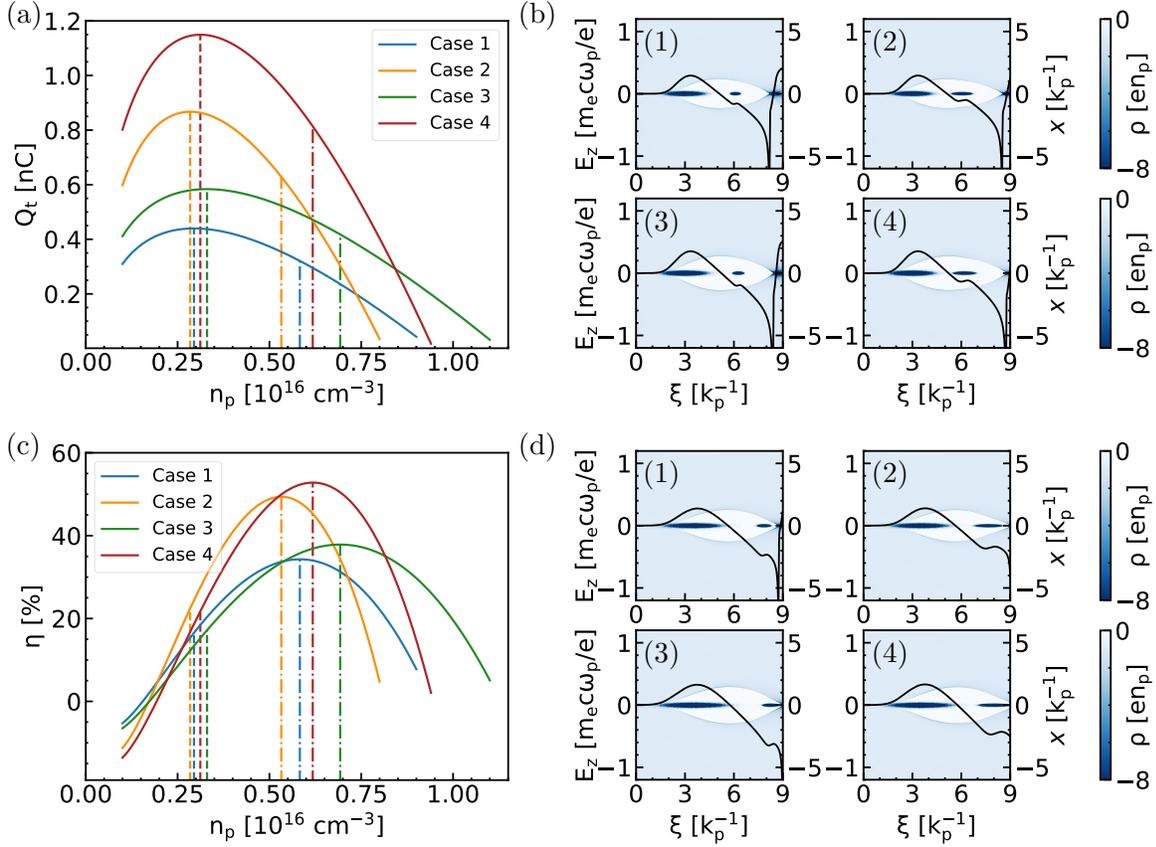
\begin{figure}[H]
    \centering
\renewcommand{\baselinestretch}{1.5}
    \begin{overpic}
        [width=0.42\textwidth,trim=0 0 70 20,clip]{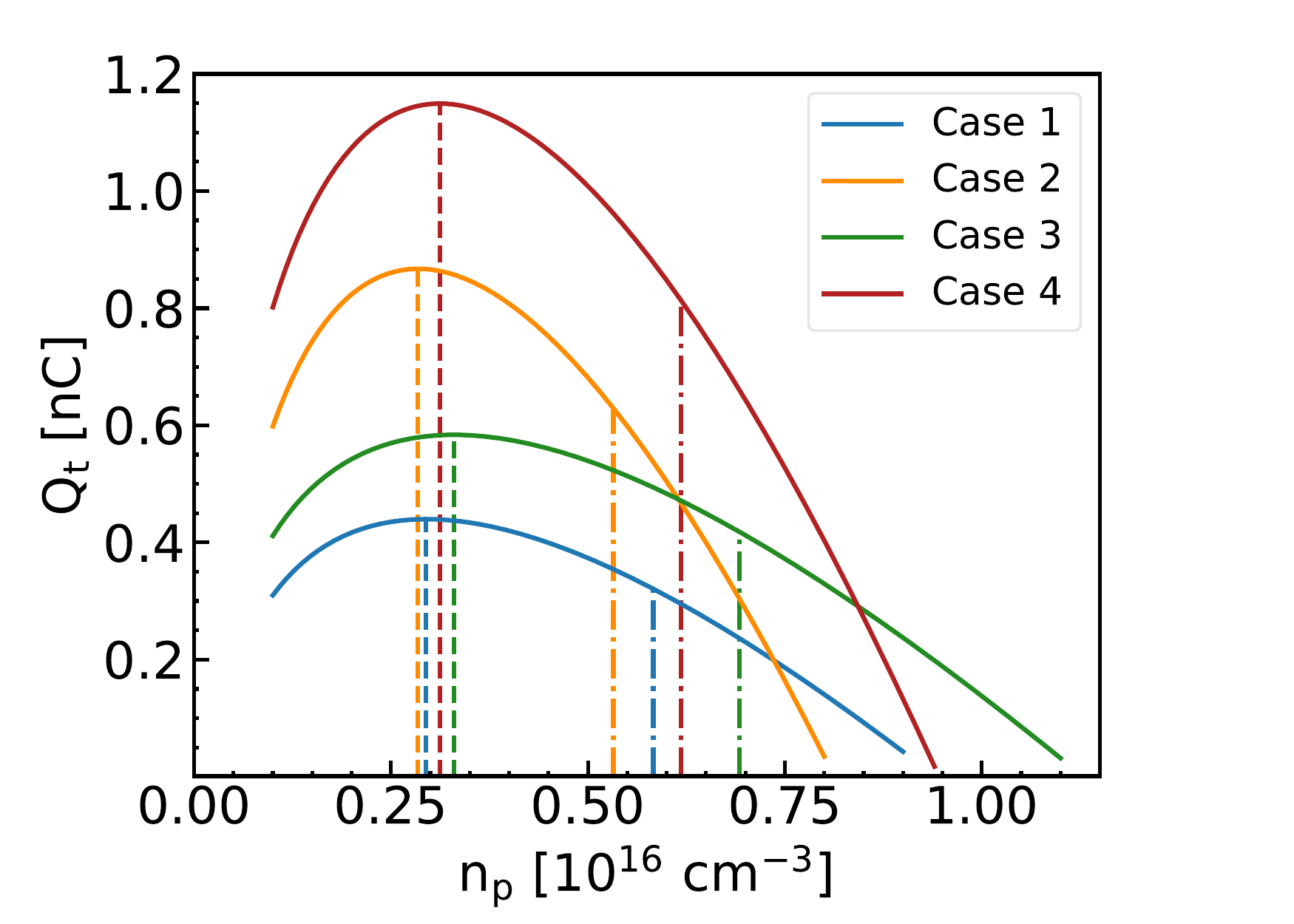}
        \put(2,76){\small{(a)}}
    \end{overpic}
    \begin{overpic}
        [width=0.56\textwidth]{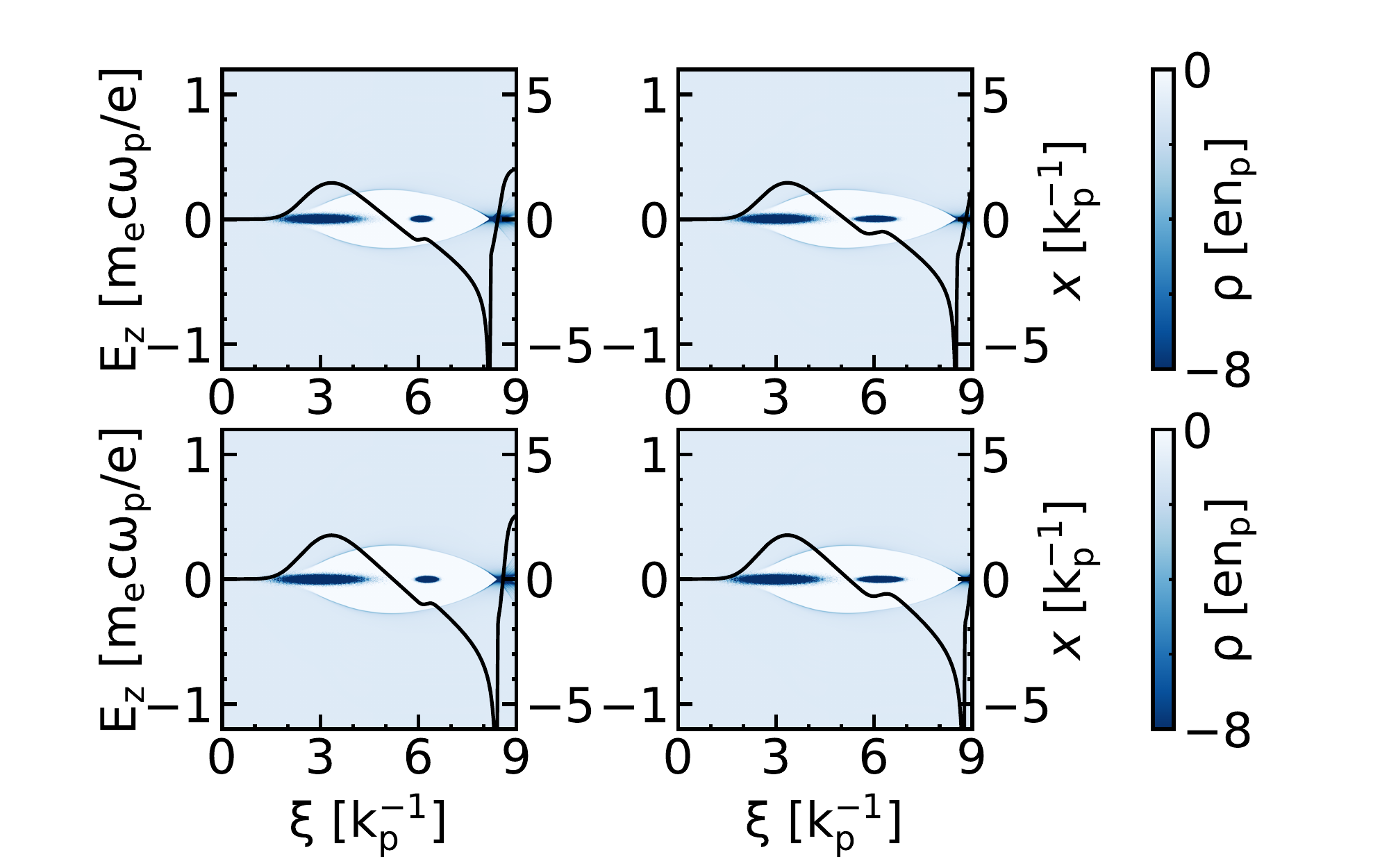}
        \put(-1,57){\small{(b)}}
        \put(17,53){\small{(1)}}
        \put(50,53){\small{(2)}}
        \put(17,27){\small{(3)}}
        \put(50,27){\small{(4)}}
    \end{overpic}
        \begin{overpic}
        [width=0.42\textwidth,trim=0 0 70 20,clip]{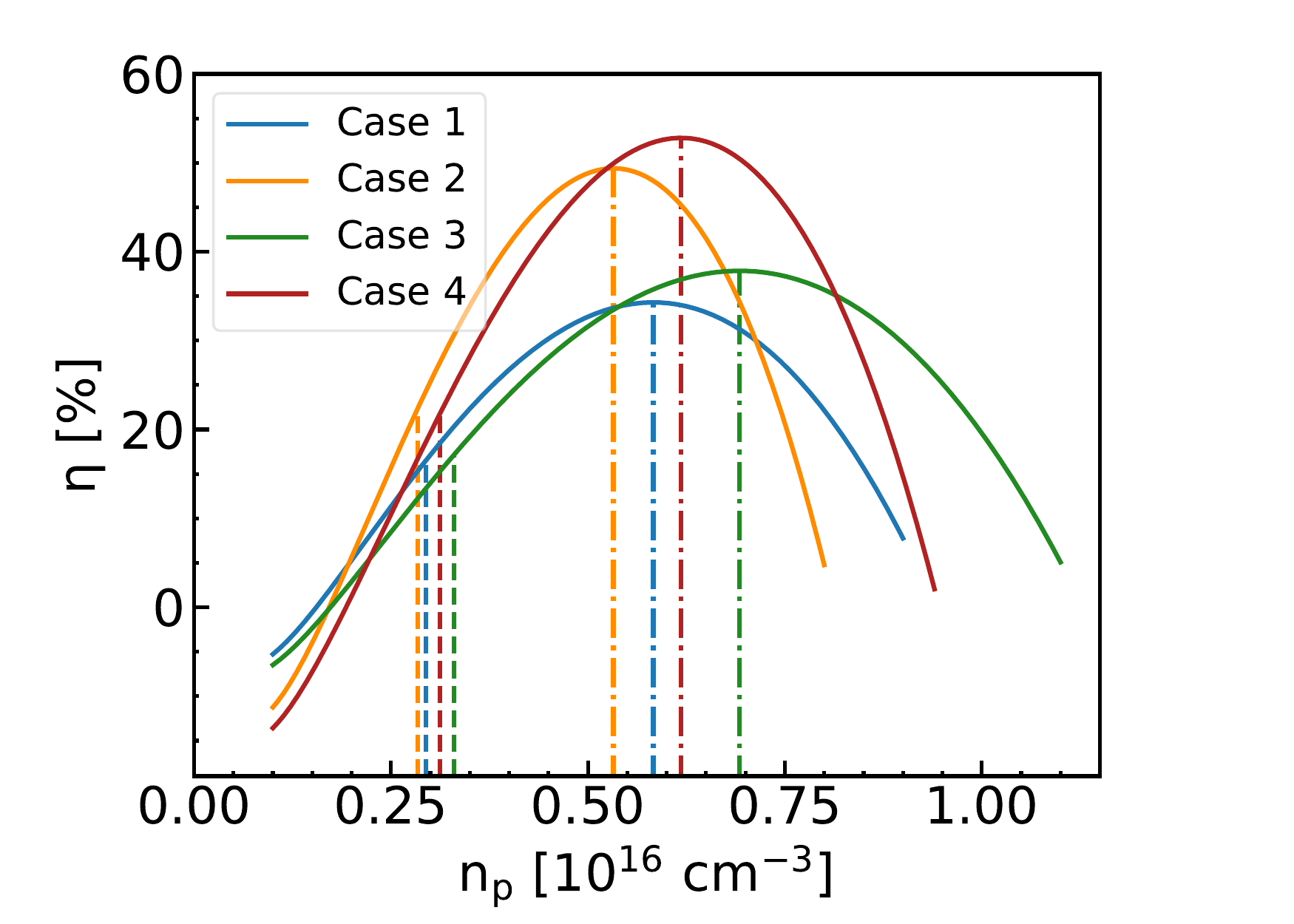}
        \put(2,76){\small{(c)}}
    \end{overpic}
    \begin{overpic}
        [width=0.56\textwidth]{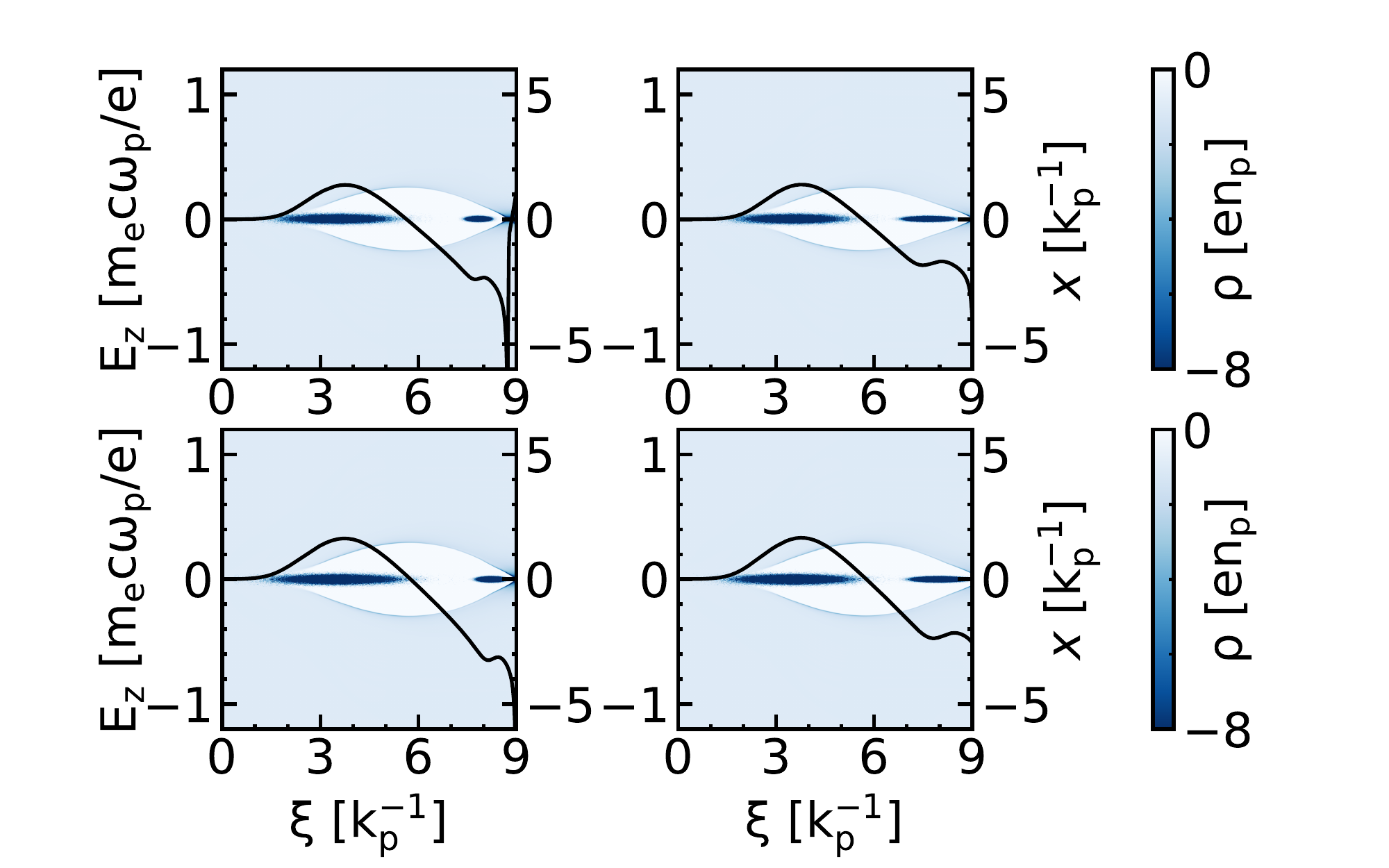}
        \put(-1,57){\small{(d)}}
        \put(17,53){\small{(1)}}
        \put(50,53){\small{(2)}}
        \put(17,27){\small{(3)}}
        \put(50,27){\small{(4)}}
    \end{overpic}
\caption{(a) $Q_t[{\rm nC}]$ versus plasma density $n_p[{\rm 10^{16} cm^{-3}}]$ under the optimal beam-loading. We have $L_d = 60 {\rm\ \mu m}$ and $l = 300
\mu m$ for these four cases. The dashed lines show the location of the optimal plasma densities for the maximum accelerated charge. (b) QuickPIC results of the plasma, beam density and the on-axis $E_z$ at the optimal densities for the maximal $Q_t$ in four different cases shown in (a). (c) $\eta$ versus plasma density $n_p[{\rm 10^{16} cm^{-3}}]$ under the optimal beam-loading. All lines have the same $Q_d, L_d, L_t$ and $l$ as those shown in (a). The dot-dashed lines show the location of the optimal plasma densities for the maximal $\eta$. (d) QuickPIC results of the plasma, beam density and the on-axis $E_z$ at the optimal densities for the maximal $\eta$ in four different cases shown in (c). QuickPIC results (1), (2), (3) and (4) in (b) and (d) correspond to the blue one, the orange one, the green one and the brown one in (a) and (c), respectively. In each simulation, we have the rms spot size of the drive beam $\sigma_{rd}[{\rm \mu m}]=5.314/{\sqrt{n_p[{\rm 10^{16} cm^{-3}}]}}$ and that of the trailing beam $\sigma_{rt}=\sigma_{rd}/2$. The simulation box has the size of $12.0\times 12.0\times 9.0$ ($x,y,\xi$) and contains $1024\times 1024\times 512$ cells.}\label{f6}
\end{figure}

The Eq.~(\ref{eq5}) shows the relation between the charge of the trailing beam and the plasma density under the optimal beam-loading. For example, when $Q_d = 1.5{\rm nC}, L_d = 60{\rm \mu m}, L_t = 12{\rm \mu m}$ and $l = 300{\rm \mu m}$, we can obtain $Q_t[{\rm nC}] = 0.1875 n_p^{\frac{3}{2}}[{\rm 10^{16} cm^{-3}}]-2.7919 n_p[{\rm 10^{16} cm^{-3}}]+2.8656 n_p^{\frac{1}{2}}[{\rm 10^{16}cm^{-3}}]-0.3230$,  which is plotted as the blue line in Fig.~\ref{f6} (a). 
The plot shows an interesting feature that the charge of the trailing beam has a maximal value when the plasma density varies under the optimal beam-loading. For this case, the trailing beam reaches its maximal charge $Q_{tmax} = 0.440{\rm nC}$ at an optimal plasma density $n_{pQ}=2.946\times {\rm 10^{15} cm^{-3}}$, which is marked as the blue dashed line in Fig.~\ref{f6} (a). We plot the plasma wake and the on-axis $E_z$ for the same case in Fig.~\ref{f6} (b)(1), and we can see that the trailing beam reaches the optimal beam-loading. In Fig.~\ref{f6} (a), we also plot Eq.~(\ref{eq5}) as three other lines with different $Q_d$ or $L_t$. Parameters for each case are listed in Table~\ref{ta5}.
\begin{table}[htb]\small
\caption{\label{ta5}
Optimal $n_p$, maximal $Q_t$ and maximal $\eta$ for cases plotted in Fig.~\ref{f6}.}
\renewcommand{\arraystretch}{1.5}
{\begin{tabular}{C{0.05\textwidth}C{0.1\textwidth}C{0.1\textwidth}C{0.15\textwidth}C{0.14\textwidth}C{0.14\textwidth}C{0.1\textwidth}C{0.16\textwidth}} \toprule
Case & $Q_d\ [\rm{nC}]$ & $L_t\ [\rm{\mu m}]$ & $n_{pQ}[{\rm 10^{15} cm^{-3}}]$ & $Q_{tmax}\ \rm{[nC]}$ & $n_{p\eta}[{\rm 10^{15} cm^{-3}}]$ & $\eta_{max}\ [\%]$ & $Q_{t}\ \rm{[nC]}$ at  $\eta_{max}$\\ \colrule
1 & 1.5 & 12 & 2.946 & 0.440 & 5.829 & 34 & 0.321 \\
2 & 1.5 & 24 & 2.843 & 0.867 & 5.323 & 49 & 0.629 \\
3 & 2 & 12 & 3.302 & 0.584 & 6.926 & 38 & 0.418 \\
4 & 2 & 24 & 3.119 & 1.149 & 6.180 & 53 & 0.814\\\botrule
\end{tabular}}
\end{table}
They all show that there is an optimal plasma density $n_{pQ}$ (marked as the dashed lines) for obtaining the maximum accelerated charge $Q_{tmax}$. The values of $n_{pQ}$ and $Q_{tmax}$ for each case are also listed in Table~\ref{ta5}. For each case, we plot the plasma wake and the on-axis $E_z$ at the optimal plasma density in Fig.~\ref{f6} (b).

Although the trailing beam reaches its maximum charge, the transformer ratio in each case is low (less than 1) as shown in Fig.~\ref{f6} (b). In other words, the acceleration efficiency is low for these cases. Actually, it is easy to find how the acceleration efficiency varies with regard to the plasma density. The acceleration efficiency can be calculated through $\eta = (Q_t[{\rm nC}]/Q_d[{\rm nC}])\cdot R$. By switching the units back to the original ones in Eq.~(\ref{eq3}) and substituting Eq.~(\ref{eq5}) into it, we can have an engineering formula of R that depends on $Q_d, L_d, L_t,l$ and $n_p$,
\begin{align}
\begin{split}
    R &= O\cdot n_p^{3}[{\rm 10^{16} cm^{-3}}]+Y\cdot n_p^{\frac{5}{2}}[{\rm 10^{16} cm^{-3}}]+Z\cdot n_p^{2}[{\rm 10^{16} cm^{-3}}]+C\cdot n_p^{\frac{3}{2}}[{\rm 10^{16} cm^{-3}}]\\
    &+K\cdot n_p[{\rm 10^{16} cm^{-3}}]+W\cdot n_p^{\frac{1}{2}}[{\rm 10^{16} cm^{-3}}]+T,\label{eq6}
\end{split}
\end{align}
where $O = m_{20}H^2/L_t^2,\ 
Y = m_{20}(2HM)/L_t^2,\ 
Z = m_{14}HL_d/L_t+m_{17}H+m_{19}Hl/L_t+m_{20}(2HP+M^2)/L_t^2,\ 
C = m_{5}H/L_t+m_{10}HQ_d/(L_dL_t)+m_{14}ML_d/L_t+m_{17}M+m_{19}Ml/L_t+m_{20}(2HS+2MP)/L_t^2,\ 
K = m_{11}L_d^2+m_{12}L_dL_t+m_{13}L_dl+m_{15}L_t^2+m_{16}L_tl+m_{18}l^2+m_{5}M/L_t+m_{10}MQ_d/(L_dL_t)+m_{14}PL_d/L_t+m_{17}P+m_{19}Pl/L_t+m_{20}(2MS+P^2)/L_t^2,\ 
W = m_2L_d+m_3L_t+m_4l+m_7Q_d+m_8Q_dL_t/L_d+m_9Q_dl/L_d+m_{14}SL_d/L_t+m_{17}S+m_{19}Sl/L_t+m_{5}P/L_t+m_{10}PQ_d/(L_dL_t)+m_{20}(2SP)/L_t^2$ and $T = m_1Q_d/L_d+m_5S/L_t+m_6Q_d^2/L_d^2+m_{10}SQ_d/(L_dL_t)+m_{20}S^2/L_t^2+m_0$.
The coefficients are given in Table~\ref{ta6}.
\begin{table}[htb]\small
\caption{\label{ta6}
Fitting coefficients in Eq.~(\ref{eq6}).}
\renewcommand{\arraystretch}{1.5}
\begin{ruledtabular}
\begin{tabular}{lllll}
$m_0$=-1.453 & $m_1$=4.489 & $m_2$=5.980$\times 10^{-3}$ & $m_3$=5.804$\times 10^{-3}$ & $m_4$=1.363$\times 10^{-2}$\\
$m_5$=-1.186$\times 10^{1}$ & $m_6$=5.355 & $m_7$=1.283$\times 10^{-1}$ & $m_8$=1.093$\times 10^{-1}$ & $m_9$=-2.825$\times 10^{-2}$\\
$m_{10}$=-5.438 & $m_{11}$=-9.840$\times 10^{-5}$ & $m_{12}$=-1.745$\times 10^{-4}$ & $m_{13}$=8.639$\times 10^{-5}$ & $m_{14}$=-9.720$\times 10^{-2}$\\
$m_{15}$=5.777$\times 10^{-4}$ & $m_{16}$=-2.269$\times 10^{-4}$ & $m_{17}$=-2.650$\times 10^{-3}$ & $m_{18}$=-6.076$\times 10^{-6}$ & $m_{19}$=3.778$\times 10^{-3}$\\
$m_{20}$=2.834$\times 10^{1}$\\
\end{tabular}
\end{ruledtabular}
\end{table}

Then by substituting Eq.~(\ref{eq5}) together with Eq.~(\ref{eq6}) into the equation of $\eta$, we can have
\begin{align}
\begin{split}
    \eta &= \frac{Q_t[{\rm nC}]}{Q_d[{\rm nC}]}\cdot R\\
         &= \frac{1}{Q_d[{\rm nC}]}\cdot(X_1\cdot n_p^{\frac{9}{2}}[{\rm 10^{16} cm^{-3}}]+X_2\cdot n_p^{4}[{\rm 10^{16} cm^{-3}}]+X_3\cdot n_p^{\frac{7}{2}}[{\rm 10^{16} cm^{-3}}]\\
         &+X_4\cdot n_p^{3}[{\rm 10^{16} cm^{-3}}]+X_5\cdot n_p^{\frac{5}{2}}[{\rm 10^{16} cm^{-3}}]+X_6\cdot n_p^{2}[{\rm 10^{16} cm^{-3}}]+X_7\cdot n_p^{\frac{3}{2}}[{\rm 10^{16} cm^{-3}}]\\
         &+X_8\cdot n_p[{\rm 10^{16} cm^{-3}}]+X_9\cdot n_p^{\frac{1}{2}}[{\rm 10^{16} cm^{-3}}]+X_{10}),\label{eq7}
\end{split}
\end{align}
where 
$X_1 = OH,\ X_2 = (OM+YH),\ X_3 =(OP+YM+ZH),\ X_4 =(OS+YP+ZM+CH),\ X_5 =(YS+ZP+CM+KH),\ X_6 = (ZS+CP+KM+WH),\ X_7 =(CS+KP+WM+TH) ,\ X_8 =(KS+WP+TM) ,\ X_9 =(WS+TP)$ and $X_{10} =TS $. 

In Fig.~\ref{f6} (c), we plot $\eta$ versus $n_p$ with four sets of $Q_d, L_d, L_t$ and $l$, which are the same as those in Fig.~\ref{f6} (a). There is also an optimal plasma density (marked as the dot-dashed lines) for obtaining the maximum $\eta$ under the optimal beam-loading. Note that $\eta$ becomes negative at lower $n_p$ because the beam separation is so small that the trailing beam is located in the decelerating phase in the plasma wake. Table~\ref{ta5} also lists the optimal plasma density $n_{p\eta}$ for the maximum acceleration efficiency $\eta_{max}$ and $Q_t$ at $\eta_{max}$. Fig.~\ref{f6} (d) shows the plasma wake and the on-axis $E_z$ at the optimal $n_p$ for the maximum acceleration efficiency for each case in Fig.~\ref{f6} (c). We can see that trailing beams are all located at the back of the bubble, which ensures that the transformer ratio is close to or larger than 1. By comparing Fig.~\ref{f6} (a) and (c), we can see that the optimal plasma densities for maximum accelerated charge and maximum acceleration efficiency are usually different. This means that for given $Q_d, L_d, L_t$ and $l$, we have to make a compromise between having the maximum accelerated charge and having the maximum acceleration efficiency when choosing the plasma density. In order to do that, for example, we can choose the value in the middle of two optimal plasma densities. In addition, the curves shown in Fig.~\ref{f6} (a) also indicate that the optimal beam-loading condition cannot hold for fixed beam parameters at different plasma densities. Therefore, additional energy spread will be induced in the region where the plasma density varies (e.g. the plasma density ramps).

\section{Conclusion}

By using the BFGS optimization method and the quasi-static code QuickPIC, we obtain a large amount of optimal beam-loading cases of two-bunch PWFA in a wide parameter range. Then we derive two fitting formulas from these data by using the polynomial regression with 10-fold cross-validation method. One fitting formula can find the optimal $\Lambda_t$ under the optimal beam-loading condition with given $\Lambda_d, \sigma_{zd}$, $\sigma_{zt}$ and $d$. The other one can find the transformer ratio with given $\Lambda_d, \sigma_{zd}, \Lambda_t, \sigma_{zt}$ and $d$ under the optimal beam-loading condition. We use the normalized units in these two fitting formulas that makes them not have the dependency of the plasma density. One can easily transform the fitting formula into an engineering equation that has the plasma density as a variable (shown as Eq.~(\ref{eq8}) and Eq.~(\ref{eq6})). The fitting formulas agree with the simulation results very well. It is a very efficient tool for obtaining the optimal beam-loading parameters when designing a PWFA stage using two tri-Gaussian electron beams in the blowout regime. We also test the fitting formulas with trailing beam that has a flat-top or trapezoidal longitudinal profile. The fitting formulas can still give a good estimation after the simple parameter transformation between different longitudinal profiles.

We explore new physics features of the optimal beam-loading based on the fitting formulas. One feature is that once the optimal beam-loading is reached, it is always satisfied when we increase the charges of drive beam and trailing beam at the same ratio. This ratio is dependent on the length of drive and trailing beams and the beam separation. Another physics feature is that under the optimal beam-loading condition there are two optimal plasma densities for the maximum accelerated charge and the maximum acceleration efficiency with given parameters of the drive beam, the length of the trailing beam and the beam separation. These two features provide an important guidance for the two-bunch PWFA design.

\begin{acknowledgments}
This study was supported by the National Natural Science Foundation of China (NSFC) grant No.\ 12075030, No.\ 11975252 and No.\ 11991071, Key Research Program of Frontier Sciences of Chinese Academy of Sciences grant No.\ QYZDJ-SSW-SLH004, Research Foundation of Institute of High Energy Physics of Chinese Academy of Sciences grant No.\ E05153U1, No.\ E15453U2, Y9545160U2 and Y9291305U2, US DOE grant No.\ DE-SC0010064, DOE SciDAC through FNAL Subcontract No.\ 644405 and NSF grant No.\ 1734315, No.\ 1806046 and No.\ 2108970 at UCLA.
\end{acknowledgments}

\appendix

\section{The Main Loop of Automatic Optimizations}

\begin{itemize}
\item[Step 1.] Initialize $N_s$ parameter sets ($\Lambda_d, \sigma_{zd}, \sigma_{zt}, d$) and $j = 1$.
\item[Step 2.] Terminate if $j>N_s$.    
\item[Step 3.]  Get the jth set of parameter ($\Lambda_{dj}, \sigma_{zdj}, \sigma_{ztj}, d_{dtj}$) and set up the input parameters for QuickPIC simulation.
\item[Step 4.]  If the trailing beam ($\pm3\sigma_{zt}$) locates in the accelerating phase, call the BFGS subroutine to calculate the optimal $\Lambda_{tj}$, which requires to call QuickPIC to calculate the value of the objective function Eq.~(\ref{eq1}). Otherwise, go to Step 6. 
\item[Step 5.] Dump the results.    
\item[Step 6.] Set $j = j + 1$ and go to Step 2.    

\end{itemize}

\section{Simulation Settings for Automatic Optimizations}
In the process of automatic optimizations, we set the center of the drive beam $C_d$ as
\begin{widetext}
\begin{equation}
    C_d = 
    \begin{cases}
    round\ (3.5\sigma_{zd},2), &\sigma_{zd}\geq 0.3 \ {\rm{and}}\ 4\sigma_{zt}<3.5\sigma_{zd}+d,\\
    ceil\ (4\sigma_{zt}-d), &4\sigma_{zt}\geq 1+d\ {\rm{or}}\ 4\sigma_{zt}\geq 3.5\sigma_{zd}+d,\\
    1, &\sigma_{zd}<0.3\ {\rm{and}}\ 4\sigma_{zt}<1+d,\nonumber
    \end{cases}
\end{equation}
\end{widetext}
where function $round\ (x,m)$ is to round $x$ to a specific precision $m$ in decimal digits and function $ceil\ (x)$ is to return the ceiling of $x$ as an integral. Subsequently, the center of the trailing beam $C_t$ is set to
\begin{equation}
    C_t = C_d+d\nonumber.
\end{equation}
 Following this, the length of box in the longitudinal direction $\rm{box_z}$ can be set to
\begin{gather}
    {\rm box_z} = max\ \lbrace 4, round\ (max\ (C_t+5\sigma_{zt},C_d+3.5\sigma_{zd}),2)\rbrace \nonumber,
\end{gather}
and the cell number in the longitudinal direction is
\begin{equation}
N_z = 
    \begin{cases}
     512, &{\rm box_z}\leq 15,\\
     1024, &{\rm box_z}>15\nonumber.
    \end{cases}
\end{equation}
In addition, the length of box in the transverse direction $\rm box_{x/y}$ can be set to
\begin{equation}
{\rm box_{x/y}} = 
    \begin{cases}
     4, &\Lambda_d\leq0.05,\\
     8R_{bmax}, &\Lambda_d>0.05.\nonumber
    \end{cases}
\end{equation}
In order to well resolve the maximal bubble radius, the cell number in the transverse direction is set to
\begin{equation}
N_{x/y} = 
    \begin{cases}
    512, &\Lambda_d\leq 0.2,\\
    1024,&\Lambda_d>0.2.\\
    \end{cases}\nonumber
\end{equation}

\section{The engineering equation for $\Lambda_t$ with the plasma density as a variable}
We can convert Eq.~(\ref{eq2}) into an equation that has the plasma density as an additional variable,
\begin{widetext}
\begin{align}
\begin{split}
    \Lambda_t = I\cdot n_p^{\frac{3}{2}}[{\rm 10^{16} cm^{-3}}]+J\cdot n_p[{\rm 10^{16}cm^{-3}}]+U\cdot n_p^{\frac{1}{2}}[{\rm 10^{16}cm^{-3}}]+V,\label{eq8}
\end{split}
\end{align}
\end{widetext}
where $I = s_{25}L_d^3+s_{26}L_d^2L_t+s_{27}L_d^2l+s_{28}L_dL_t^2+s_{29}L_dL_tl+s_{30}L_dl^2+s_{31}L_t^3+s_{32}L_t^2l+s_{33}L_t l^2+s_{34}l^3,\ 
J = s_9L_d^2+s_{10}L_dL_t+s_{11}L_dl+s_{12}L_t^2+s_{13}L_tl+h_{14}l^2+s_{19}\Lambda_dL_d^2+s_{20}\Lambda_dL_dL_t+s_{21}\Lambda_dL_dl+s_{22}\Lambda_dL_t^2+s_{23}\Lambda_dL_tl+s_{24}\Lambda_dl^2,\ 
U = s_2L_d+s_3L_t+s_4l+s_6\Lambda_dL_d+s_7\Lambda_dL_t+s_8\Lambda_dl+s_{16}\Lambda_d^2L_d+s_{17}\Lambda_d^2L_t+s_{18}\Lambda_d^2l,\
V = s_1\Lambda_d+s_5\Lambda_d^2+s_{15}\Lambda_d^3+s_0$ and $L_d, L_t$ and $l$ are in the unit of $\mu m$.
The coefficients are given in Table~\ref{ta7}.
\begin{table}[H]\small
\caption{\label{ta7}
Fitting coefficients in Eq.~(\ref{eq8}).}
\renewcommand{\arraystretch}{1.5}
\setlength{\tabcolsep}{0.3mm}{
\begin{ruledtabular}
\begin{tabular}{lllll}
$s_0$=-5.014$\times 10^{-1}$ & $s_1$=3.658$\times 10^{-1}$ & $s_2$=1.716$\times 10^{-2}$ & $s_3$=-2.038$\times 10^{-2}$ & $s_4$=5.763$\times 10^{-3}$\\
$s_5$=-3.754$\times 10^{-2}$ & $s_6$=4.411$\times 10^{-2}$ & $s_7$=2.410$\times 10^{-3}$ & $s_8$=-9.462$\times 10^{-4}$ & $s_9$=-2.527$\times 10^{-4}$\\
$s_{10}$=-6.781$\times 10^{-5}$ & $s_{11}$=-4.661$\times 10^{-5}$ & $s_{12}$=-7.673$\times 10^{-4}$ & $s_{13}$=3.663$\times 10^{-4}$ & $s_{14}$=-2.694$\times 10^{-5}$\\
$s_{15}$=-2.391$\times 10^{-3}$ & $s_{16}$=-1.425$\times 10^{-3}$ & $s_{17}$=4.970$\times 10^{-4}$ & $s_{18}$=2.182$\times 10^{-4}$ & $s_{19}$=-3.055$\times 10^{-4}$\\ 
$s_{20}$=-8.582$\times 10^{-5}$ & $s_{21}$=3.410$\times 10^{-5}$ & $s_{22}$=1.372$\times 10^{-4}$ & $s_{23}$=-2.527$\times 10^{-5}$ & $s_{24}$=-1.084$\times 10^{-6}$\\ 
$s_{25}$=8.251$\times 10^{-7}$ & $s_{26}$=2.500$\times 10^{-7}$ & $s_{27}$=5.101$\times 10^{-7}$ & $s_{28}$=3.463$\times 10^{-6}$ & $s_{29}$=-1.723$\times 10^{-7}$\\
$s_{30}$=-4.045$\times 10^{-8}$ & $s_{31}$=-1.910$\times 10^{-5}$ & $s_{32}$=8.201$\times 10^{-6}$ & $s_{33}$=-1.682$\times 10^{-6}$ & $s_{34}$=4.448$\times 10^{-8}$\\
\end{tabular}
\end{ruledtabular}}
\end{table}

\bibliography{apssamp}

\end{document}